\documentclass[aps, prfluids, showpacs, showkeys, notitlepage, amsmath, amssymb, floatfix, longbibliography, 12pt, tightenlines]{revtex4-2}
\usepackage[export]{adjustbox}
\usepackage{svg}
\usepackage{float}
\usepackage{amsmath}
\usepackage{amssymb}
\usepackage{multirow}
\usepackage[font=small, labelfont=bf, justification=justified]{caption}
\usepackage{subfig}
\usepackage{graphicx}% Include figure files
\usepackage{dcolumn}% Align table columns on decimal point
\usepackage{bm}% bold math
\usepackage{hyperref}% add hypertext capabilities
\usepackage[normalem]{ulem}

\DeclareMathAlphabet{\altmathcal}{OMS}{cmsy}{m}{n}

\begin{document}

\title{Extreme vertical drafts as drivers of Lagrangian dispersion in stably stratified turbulent flows}
\author{Christian Reartes}
\affiliation{Universidad de Buenos Aires, Facultad de Ciencias Exactas y Naturales, Departamento de Física, Ciudad Universitaria, 1428 Buenos Aires, Argentina}
\affiliation{CONICET - Universidad de Buenos Aires, Instituto de F\'{\i}sica Interdisciplinaria y Aplicada (INFINA), Ciudad Universitaria, 1428 Buenos Aires, Argentina}
\author{Pablo D.~Mininni}
\affiliation{Universidad de Buenos Aires, Facultad de Ciencias Exactas y Naturales, Departamento de Física, Ciudad Universitaria, 1428 Buenos Aires, Argentina}
\affiliation{CONICET - Universidad de Buenos Aires, Instituto de F\'{\i}sica Interdisciplinaria y Aplicada (INFINA), Ciudad Universitaria, 1428 Buenos Aires, Argentina}
\author{Raffaele Marino}
\affiliation{CNRS, \'Ecole Centrale de Lyon, INSA Lyon, Universit\'e Claude Bernard Lyon 1, LMFA, UMR5509, F-69134 \'Ecully, France}

\begin{abstract}

The dispersion of Lagrangian particle pairs is a fundamental process in turbulence, with implications for mixing, transport, and the statistical properties of particles in geophysical and environmental flows. While classical theories describe pair dispersion through scaling laws related to energy cascades, extreme events in turbulent flows can significantly alter these dynamics. This is especially important in stratified flows, where intermittency manifests itself also as strong updrafts and downdrafts. In this study, we investigate the influence of extreme events on the relative dispersion of particle pairs in stably stratified turbulence. Using numerical simulations we analyze the statistical properties of pair separation across different regimes, and quantify deviations from classical Richardson scaling. Our results highlight the role of extreme drafts in accelerating dispersion. These findings have important implications for turbulent mixing in natural systems, including atmospheric and oceanic flows, as well as applications in cloud microphysics and pollutant transport.
\end{abstract}

\maketitle

\section{Introduction}

Stratified turbulent flows are ubiquitous in geophysical systems, ranging from the atmosphere to the deep ocean, and play a central role in regulating heat transport, mass transfer, and mixing processes \cite{Winters_1995, Metais_1996, Cambon_1999, Smith_2002b, Riley2008, gregg2018mixing, Smyth_2019}. Under stable stratification, buoyancy forces suppress vertical motions, giving rise to a complex interplay between internal gravity waves and turbulent eddies \cite{waite_bartello_2004, lindborg2006energy}. Although stratification generally inhibits turbulence and vertical mixing, numerous studies have reported the spontaneous occurrence of extreme events—localized vertical drafts and intense bursts of dissipation—whose magnitude and intermittency exceed Gaussian predictions, even beyond classical corrections for intermittency in homogeneous and isotropic turbulence \cite{Rorai_2014, Feraco_2018, Sujovolsky2019, Marino2022}.

Although rare and localized, extreme events can dominate the energy budget and strongly affect mixing efficiency, as demonstrated by numerical and theoretical studies \cite{Feraco_2018, Marino2022}. Comparable results have been reported across a range of Prandtl numbers \cite{petropoulos2024extreme}, thereby broadening the conditions under which such events arise. Their emergence has been associated with nonlinear interactions among internal wave dynamics, shear instabilities, and small-scale turbulence \cite{Sujovolsky2019, feraco2021connecting, Sujovolsky2021}. Intense vertical drafts and temperature bursts have been shown to develop spontaneously in stratified flows, triggering local energy cascades and markedly enhancing dissipation \cite{Marino2022, Foldes2025}. This link between extreme events and elevated energy dissipation is further supported by oceanic models and atmospheric observations \cite{Pearson_2018, chau2021extreme, rodriguezimazio2023}.

Beyond their energetic role, extreme events have significant consequences for tracer and particle transport. In particular, Lagrangian pair dispersion is highly sensitive to strong vertical motions \cite{buaria2020single}. Atmospheric and oceanic observations indicate that mixing often occurs within intermittent turbulent patches surrounded by quiescent regions \cite{Pearson_2018, Marino2022, rodriguezimazio2023, Lefauve2024}. Such spatial heterogeneity challenges classical turbulence models and complicates the parameterization of mixing and particle transport in geophysical simulations \cite{BeronVera2020, Reartes2024}.

Although recent studies have advanced our understanding of large-scale intermittency and non-Gaussian statistics in stably stratified flows \cite{Feraco_2018, Sujovolsky2019, Marino2022, petropoulos2024extreme}, key questions remain. In particular, the spatial organization, frequency, mixing efficiency, and dynamical impact of extreme events are not yet fully understood. Furthermore, their influence on inertial particle dynamics and on the possible emergence of anomalous dispersion regimes remains largely unexplored \cite{Sozza2016, Reartes2023}. Addressing this gap is especially relevant for applications such as cloud microphysics, pollutant dispersion, and mass transport in the ocean \cite{AcevedoTrejos2015, Squires_1995, BeronVera2020}.

In this work we address these questions using direct numerical simulations (DNSs) of stably stratified turbulence (SST), combining a Lagrangian framework with statistical diagnostics to detect and characterize extreme vertical velocity events. By classifying particles according to their exposure to strong vertical drafts, we examine their pair dispersion, compute finite-time Lyapunov exponents (FTLEs), and evaluate enstrophy along their trajectories. Our results show that extreme events establish a conditional transport regime in which a subset of particles experiences enhanced vertical dispersion, closely associated with coherent Lagrangian structures and regions of intense energy dissipation. These deviations from classical dispersion laws demonstrate that extreme events in SST fundamentally reshape mixing dynamics. To illustrate the role of extreme events in Lagrangian transport, the Supplementary Material \cite{sm} includes a video showing the evolution of two tracer particle pairs in the same stably stratified turbulent flow. Both pairs are initialized with identical separation and evolve simultaneously under identical flow conditions. Yet, while one pair traverses a region with extreme vertical drafts, the other remains in a quiescent zone. The contrast is striking: the pair experiencing extreme events undergoes rapid, explosive dispersion, whereas the other stays closely bound. This direct comparison highlights a central message of this study---extreme events can produce fundamentally different Lagrangian trajectories for fluid elements governed by the same equations, thereby underscoring the impact of large-scale intermittency on mixing and transport.

While the primary focus of this study is the Lagrangian dynamics of tracers in stratified turbulence, we also present a brief analysis of neutrally buoyant particles. In stratified environments, such particles are vertically confined near isopycnal layers by the buoyancy force, leading to constrained vertical motion and complex transport dynamics \cite{van_aartrijk_2010, Sozza2016, Reartes2023}. Despite being small, they can display behavior that differs markedly from that of fluid elements, including vertical clustering and sensitivity to turbulence-induced fluctuations \cite{Reartes2023}. The Maxey–Riley–Gatignol equation, under the Boussinesq approximation, provides a tractable model for their dynamics in this regime \cite{Maxey_Riley_1983, Gatignol1983}. Previous studies have shown that such particles tend to accumulate in regions of low vorticity in stably stratified flows, with confinement levels strongly controlled by the Stokes and Froude numbers \cite{Reartes2023}. Here we investigate whether extreme vertical drafts---identified as Lagrangian extreme events---can disrupt this confinement and promote enhanced vertical dispersion. This analysis complements our broader effort to quantify the role of extreme events in turbulent mixing.

This study focuses on the Lagrangian manifestation of extreme events in SST, introducing a statistical framework to identify such events, quantify their contribution to mixing, and relate them to coherent structures. Combined with prior studies of large-scale intermittency in SST \cite{Feraco_2018, Smyth_2019, feraco2021connecting, Marino2022}, indirect assessments of stratified mixing \cite{petropoulos2024extreme}, and geophysical observations of patchy mixing in the atmosphere and oceans \cite{Pearson_2018, chau2021extreme, rodriguezimazio2023}, our findings highlight the central role of extreme events in shaping dispersion processes.

\section{Numerical simulations\label{sec:setup}}

In this study, we numerically solve the incompressible Boussinesq equations for the velocity field ${\bf u}$ and for mass density fluctuations $\rho'$,
\begin{equation}
\partial_t {\bf u} + {\bf u} \cdot \boldsymbol{\nabla} {\bf u} = - \boldsymbol{\nabla} \left(p/\rho_0\right) - \left(g/\rho_0 \right)\rho' \hat{z} + \nu \nabla^2 {\bf u} + {\bf f},
\end{equation}
\begin{equation}
\partial_t \rho' + {\bf u} \cdot \boldsymbol{\nabla} \rho' = \left(\rho_0 N^2/g\right) {\bf u} \cdot \hat{z} + \kappa \nabla^2 \rho',
\end{equation}
\begin{equation}
\nabla \cdot {\bf u} = 0,
\end{equation}
where $p$ denotes the correction to the hydrostatic pressure, $\nu$ the kinematic viscosity, ${\bf f}$ an external mechanical forcing, $N$ the Brunt–Väisälä frequency (which, under this approximation, sets the stratification), and $\kappa$ the diffusivity. In terms of the background density gradient, the Brunt–Väisälä frequency is defined as $N^2 = -(g/\rho_0)(d\bar{\rho}/dz)$, where $d\bar{\rho}/dz$ is the imposed linear background stratification and $\rho_0$ the mean fluid density.

We rescale the density fluctuations in velocity units by defining $\zeta = g \rho' / (\rho_0 N)$. All quantities are then nondimensionalized using a characteristic length scale $L_0$ and velocity scale $U_0$ in the domain, leading to,
\begin{equation}
\partial_t {\bf u} + {\bf u} \cdot \boldsymbol{\nabla} {\bf u} = - \boldsymbol{\nabla} \left(p/\rho_0\right) - N \zeta \hat{z} + \nu \nabla^2 {\bf u} + {\bf f},
\label{bou1}
\end{equation}
\begin{equation}
\partial_t \zeta + {\bf u} \cdot \boldsymbol{\nabla} \zeta = N {\bf u} \cdot \hat{z} + \kappa \nabla^2 \zeta.
\label{bou2}
\end{equation}

Equations~(\ref{bou1}) and (\ref{bou2}) are governed by two dimensionless control parameters: the Reynolds number and the Froude number,
\begin{equation}
\textrm{Re} = \frac{LU}{\nu}, \quad \textrm{Fr} = \frac{U}{LN},
\label{eq:Re_Fr}
\end{equation}
where $L = \pi/(2u'^2) \int E(k)/k , dk$ is the characteristic Eulerian integral scale, $U = \langle |{\bf u}|^2 \rangle^{1/2}$ the r.m.s.~velocity, $E(k)$ the isotropic kinetic energy spectrum, and $u'^2 = U^2/3$. Based on these parameters, the buoyancy Reynolds number is defined as,
\begin{equation}
\textrm{Rb} = \textrm{Re} \textrm{Fr}^2,
\end{equation}
which estimates the turbulence strength at the buoyancy scale, $L_b = U/N$, and plays a central role in characterizing flow dynamics. For $\textrm{Rb} \gg 1$ strong stratified turbulence can develop, whereas for $\textrm{Rb} \ll 1$ turbulent motions are heavily damped by viscosity. Although geophysical flows typically exhibit large buoyancy Reynolds numbers (e.g., observations in the ocean thermocline suggest $\textrm{Rb} \approx 10^2$–$10^3$ \cite{moum_1996}), computational constraints limit the values of $\textrm{Rb}$ that can be realistically simulated. Previous studies have shown that $\textrm{Rb} > 10$ is sufficient to sustain some small-scale turbulence in the flow \cite{Ivey2008, Pouquet2018}.

The Ozmidov scale, $L_{oz} = 2\pi/k_{oz}$ with $k_{oz} = \sqrt{N^3/\varepsilon}$, also plays a critical role in the dynamics. At scales much smaller than $L_{oz}$, the flow is expected to recover isotropy. When $\textrm{Rb} > 1$, the Ozmidov scale exceeds the Kolmogorov scale $\eta$, and quasi-isotropic turbulent transport can be anticipated at the smallest dynamical scales of the flow.

In this study, the choice of Froude numbers for the numerical simulations was informed by previous work identifying an optimal range for the occurrence of extreme events in stratified turbulence. In particular, Ref.~\cite{Feraco_2018} showed that, in stably stratified turbulent flows, large-scale intermittency intensifies within the range $0.05 < \textrm{Fr} < 0.3$, with a maximum around $\textrm{Fr} \approx 0.08$. In this regime, vertical velocity and potential temperature fluctuations display fat-tailed probability distributions, indicating the emergence of extreme events localized in space and time. The analysis in \cite{Feraco_2018} further revealed that the Lagrangian vertical velocity kurtosis also attains its maximum at $\textrm{Fr} \approx 0.08$. Since values of $\textrm{Fr}$ of $\mathcal{O}(10^{-2})$ are also typical of atmospheric and oceanic systems, we consider simulations in this range to capture the influence of extreme events on particle dispersion in geophysical contexts.

\begin{table*}
	\centering
	\caption{\label{tab_fluid_exev} Relevant parameters of the fluid simulations. $n_x$, $n_y$, and $n_z$ are the grid dimensions; $NT_0$ is the Brunt-Väisälä frequency in units of $T_0^{-1} = U_0/L_0$; $\textrm{Fr}$ is the Froude number, $\textrm{Re}$ the Reynolds number, and $\textrm{Rb}$ the buoyancy Reynolds number. $L$ is the integral length scale of the flow, $\eta$ the Kolmogorov scale, $L_b$ the buoyancy length, and $L_{Oz}$ the Ozmidov scale. All lengths are in units of the reference length $L_0$.}
    \begin{ruledtabular}
	\begin{tabular}{ccccccccccc}
		Run & $n_x=n_y$ & $n_z$ & $NT_0$ & $\textrm{Fr}$ & $\textrm{Re}$ & $\textrm{Rb}$ & $L/L_0$ & $\eta/L_0$ & $L_b/L_0$ & $L_{Oz}/L_0$ \\ \hline
		TG  & 768 & 192 & 8  & 0.1 & 2600 & 24 & 1.05 & 0.005 & 0.63 & 0.17\\
		RND & 1024 & 1024 & 5  & 0.05 & 7200 & 18 & 2.50 & 0.006 & 0.75 & 0.05
	\end{tabular}
    \end{ruledtabular}
\end{table*}
\begin{table*}
\caption{\label{tablaga} Relevant parameters of the inertial particles in the TG simulations. $\textrm{St}$ is the Stokes number, $\tau_p/T_0$ is the Stokes time in units of $T_0$, $a/\eta$ is the particle radius in units of the Kolmogorov scale, and $\textrm{Re}_p$ lists the corresponding particles Reynolds numbers.}
\begin{ruledtabular}
\begin{tabular}{ccccc}
Label & $\textrm{St}$ & $\tau_p/T_0$ & $a/\eta$ & $\textrm{Re}_p$\\
\hline
$St03$  & 0.3 & 0.024 & 1.08 & 0.2\\ 
$St1$   & 1   & 0.076 & 1.85 & 0.5\\ 
$St3$   & 3   & 0.235 & 3.38 & 2.7
%$St6$   & 6   & 0.470 & 4.28 & - & 7.6 & -
\label{tab_parts}
\end{tabular}
\end{ruledtabular}
\end{table*}

We performed two numerical simulations with different spatial resolutions, domain geometries, and forcing mechanisms (see Table~\ref{tab_fluid_exev}). In the first case, labeled TG, turbulence is sustained by Taylor–Green (TG) forcing \cite{clark_di_leoni_2015}, with a resolution of $N_x = N_y = 768$ and $N_z = 192$ grid points. The domain extends over $L_x = L_y = 2\pi L_0$ in the horizontal directions and $L_z = H = \pi L_0 / 2$ in the vertical, where $L_0$ is the reference length. The TG forcing is a two-component velocity forcing that produces large-scale, counter-rotating vortex pairs perpendicular to the stratification, separated by a shear layer. Its form is given by,
\begin{equation}
    {\bf f} = f_{0} \left[ \sin(k_{f}x)\cos(k_{f}y)\cos(k_{f}z) \hat{\bf x} - \cos(k_{f}x)\sin(k_{f}y)\cos(k_{f}z) \hat{\bf y} \right],
\end{equation}
where $f_0$ is the forcing amplitude and $k_f = 1/L_0$ the forcing wavenumber. This flow configuration has been widely employed to investigate stratified turbulence \cite{riley_2003, Sujo_2018}. In the stratified case, it generates a balanced large-scale circulation with vertically sheared horizontal winds (VSHWs), i.e., a non-zero mean horizontal velocity, which in this flow is confined to the shear layer separating the TG vortices \cite{Reartes2023}. The forcing is time-independent and therefore does not introduce additional time scales into the system. The Froude number in Eq.~(\ref{eq:Re_Fr}) can then be interpreted as the ratio between the characteristic frequency of the large-scale eddies and the Brunt–Väisälä frequency.

The second simulation, labeled RND, was carried out in a cubic domain with $N_x = N_y = N_z = 1024$ grid points and physical size $L_x = L_y = L_z = 2\pi L_0$. In this case, isotropic random three-dimensional forcing was applied, following the method described in \cite{Patterson1978}. This approach is used, in our case, to generate a non-helical random forcing from a superposition of Fourier modes with random phases. Two random vector fields are generated in Fourier space with Cartesian components given by
\begin{equation}
v_i^{(1)}({\bf k}) = A(k) e^{i \phi}, \hspace{1cm} v_i^{(2)}({\bf k}) = A(k) e^{i \phi'},
\label{1d1d8}
\end{equation}
where $i = 1, 2, 3$ denotes the component, $\phi$ and $\phi'$ are random phases (different for each component and wavevector ${\bf k}$), and $A(k)$ is an amplitude function controlling the shape of the forcing spectrum. Three incompressible vector fields are then constructed as:
\begin{equation}
{\bf u}^{(1)} = \boldsymbol{\nabla} \times {\bf v}^{(1)}, \hspace{1cm} {\bf u}^{(2)} = \boldsymbol{\nabla} \times {\bf v}^{(2)},
\label{eqdkjdkm}
\end{equation}
\begin{equation}
\boldsymbol{\omega}^{(3)} = \boldsymbol{\nabla} \times \left[ {\bf u}^{(1)} \sin\alpha + {\bf u}^{(2)} \cos\alpha \right].
\end{equation}
Finally, the forcing is defined as
\begin{equation}
{\bf f}({\bf k}) = {\bf u}^{(1)}({\bf k}) \sin\alpha + {\bf u}^{(2)}({\bf k}) \cos\alpha + \frac{\boldsymbol{\omega}^{(3)}({\bf k})}{k},
\end{equation}
where $\alpha$ is a parameter controlling the helicity of the forcing (the relative helicity is $h = \sin(2\alpha)$, with $h=0$ for $\alpha = 0$, and $h=1$ for $\alpha = \pi/4$). In this study, we set $\alpha = 0$ to avoid helicity injection. The random phases $\phi$ and $\phi'$ evolve in time with correlation time $\tau_f = 0.5 L_0/U_0$. The forcing is applied to all modes with wavenumber $k_f = 1/L_0$ lying on a spherical shell in Fourier space. Unlike TG forcing, RND does not generate large-scale structures in the flow. In the presence of stratification, this mechanism produces an unbalanced flow, excites strong internal gravity waves, and promotes the development of VSHWs at all heights \cite{PhysRevE.90.023018}.

The two forcing functions considered lead to markedly different flow dynamics. As already mentioned, TG forcing produces localized regions of vertical shear, constrains the growth of horizontal winds, but also promotes the formation of turbulent instabilities associated to the large-scale circulation---features useful for studying the interaction between coherent structures and particle dispersion. In contrast, RND imposes a more homogeneous energy distribution, enabling the analysis of transport in a wave- and wind-dominated regime, free from the influence of large-scale organized vortices. Together, these forcing strategies provide complementary insights into distinct transport mechanisms in SST.

For practical purposes, the Boussinesq equations, Eqs.~(\ref{bou1}) and (\ref{bou2}), were solved in a triply periodic domain using a parallelized, fully dealiased pseudo-spectral method and a second-order Runge–Kutta scheme in time \cite{mininni_hybrid_2011}. All simulations use a Prandtl number $\textrm{Pr} = \nu/\kappa = 1$, interpreted as an effective turbulent Prandtl number \cite{Li2021, basu2021}. The kinematic viscosity was chosen to ensure that the Kolmogorov scale, $\eta = (\nu^3/\varepsilon)^{1/4}$, is well resolved, where the kinetic energy dissipation rate is given by $\varepsilon = \nu \langle |\boldsymbol{\omega}|^2 \rangle$, with $\boldsymbol{\omega} = \boldsymbol{\nabla} \times {\bf u}$ the vorticity.

Lagrangian tracers were initialized randomly throughout the domain once the flows had reached a statistically steady turbulent state. Their initial velocities were set equal to the fluid velocity at the particle positions, and as Lagrangian fluid elements, they follow the local fluid velocity at all times. In each simulation listed in Table~\ref{tab_fluid_exev}, $10^6$ particles were used in the TG run and $1.5 \times 10^6$ in the RND run. The simulations were integrated for more than 120 large-eddy turnover times, providing sufficient statistics across multiple extreme events in the fully developed turbulent regime.

To complement the analysis of Lagrangian tracers, we also briefly consider neutrally buoyant inertial particles only in the TG simulation. Their dynamics are governed by a simplified form of the Maxey–Riley–Gatignol equation \cite{Maxey_Riley_1983,Gatignol1983}. These particles have the same density as the surrounding fluid, but owing to their finite response time, they do not follow the fluid velocity instantaneously \cite{Sozza2016, Reartes2023}. Their motion is described by
\begin{equation}
\Dot{\bf v} = \frac{1}{\tau_p} \left[ {\bf u}({\bf x},t) - {\bf v}(t) \right] - \frac{2}{3}N\left[N(z-z_0) - \zeta \right] \hat{z} + \frac{\textrm{D}}{\textrm{D}t} {\bf u}({\bf x},t),
\label{eq:inertial_dynamics}
\end{equation}
where, for a spherical particle of radius $a$, $\tau_p = a^2/(3\nu)$ is the particle response time and $\zeta$ the rescaled buoyancy field. The Stokes number is defined as $\textrm{St} = \tau_p/\tau_\eta$, where $\tau_\eta = (\nu/\varepsilon)^{1/2}$ is the Kolmogorov time scale. In our simulations, we restrict to buoyancy Stokes numbers $\textrm{Sb} = \tau_p N < 1$, for which the Basset–Boussinesq history term is negligible \cite{Reartes2024}. We consider three values of $\textrm{St}$, spanning weak to moderate inertia (see Table~\ref{tab_parts}). 
Each set in table \ref{tab_parts} has $10^6$ particles, which were evolved for 80 large-eddy turnover times to ensure sufficient statistical sampling. 
Particles are initialized in a horizontal slab of width $H/5$ centered at $z_0 = H/2$, corresponding to the shear layer where extreme events are most prevalent in the TG flow; this configuration maximizes the likelihood of particle–draft interactions. 
Particles are one-way coupled, do not interact with one another, and do not modify the flow. The particle Reynolds number, defined as $\textrm{Re}_p = a |{\bf u} - {\bf v}|/\nu$, remains below or of order unity in all cases. 
These particles should be considered only as test particles to measure the impact of the extreme events when particles have inertia, other cases of particles (as, e.g., heavy particles relevant for cloud physics) are left for future studies.

\section{Lagrangian definition of extreme events}

\begin{figure}
	\centering
	\includegraphics[width=1\textwidth]{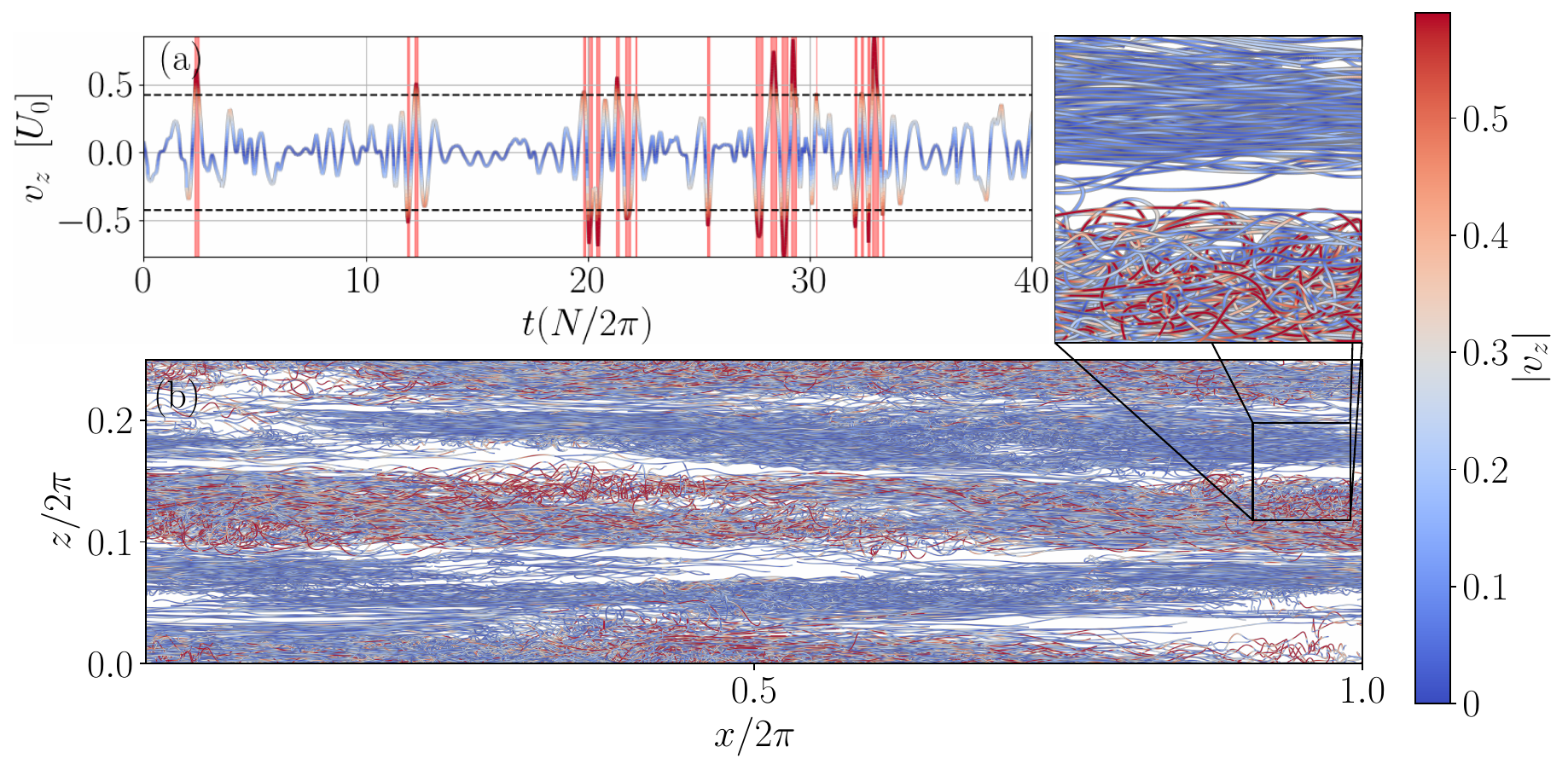}
	\caption{(a) Vertical velocity $v_z$ of a representative tracer as a function of time, normalized by the Brunt–Väisälä period. An extreme event is defined when $v_z$ exceeds $4\sigma_{v_z}$ (dashed line), where $\sigma_{v_z}$ is the instantaneous standard deviation of $v_z$ across all tracers. The cumulative time spent in extreme events is shown as red-shaded intervals. (b) Trajectories of tracers in the TG simulation on an $x$–$z$ plane during one Brunt–Väisälä period, coloured by the magnitude of the vertical velocity (from blue to red). Trajectories experiencing extreme vertical velocities appear more tangled and mixed, whereas those avoiding extreme events are more ordered. An inset provides a detail highlighting the contrast between tracers that cross extreme drafts and those that do not.}
	\label{def_t_exev}
\end{figure}

To assess the effect of extreme events on particle transport, a clear criterion is required to identify whether a particle (either a tracer or an inertial particle) experiences such an event. In this study, a particle is considered to undergo an extreme event when its vertical velocity exceeds the threshold $4\sigma_{v_z}$ at a given instant $t$, where $\sigma_{v_z}$ is the standard deviation of the instantaneous vertical velocity of all particles in the same dataset at the same time. The dimensionless cumulative time a particle spends in extreme events is then computed as
\begin{equation}
\tau_{exev} = \left( \frac{N}{2\pi} \right) \int_{|v_z| > 4 \sigma_{v_z}} dt \approx \left( \frac{N}{2\pi} \right) \sum_{i} n_i \Delta t ,
\end{equation}
i.e., the sum of the number of time steps $n_i$ spanning the time intervals during which this condition is satisfied, multiplied by the time step $\Delta t$, and normalized by the Brunt–Väisälä period (see Fig.~\ref{def_t_exev}(a) for an illustration). The rationale for adopting this threshold is illustrated in Fig.~\ref{def_t_exev}: panel (b) shows tracer trajectories in the TG simulation on an $x$–$z$ plane, coloured by their vertical velocity magnitude. Trajectories that cross regions of extreme vertical drafts appear strongly tangled and mixed, whereas those that do not experience such events remain more ordered. This clear separation supports the use of $4\sigma_{v_z}$ as an effective criterion for identifying Lagrangian extreme events.

Based on this criterion, we classify all particles into two main subsets. The first consists of particles that experience sufficiently long extreme events, defined as those with $\tau_{exev} > 1$. The second subset includes particles that remain in a “mild’’ flow, i.e., particles that never encounter an extreme event and whose vertical velocity stays below $2.5\sigma_{v_z}$ throughout their evolution. These definitions leave a third residual set of particles that do not satisfy either condition---those for which $|v_z|$ becomes large but not large enough, or not sustained for long enough, to qualify as particles going through extreme events.

This classification is motivated by the presence of invariant manifolds in the phase space of stratified turbulence---structures along which fluid elements evolve slowly, with rapid transitions between them \cite{Sujovolsky2019, Sujovolsky2021}. Previous studies have shown that fluid elements in stratified turbulence alternate between two main dynamical states: one dominated by internal waves, where vertical transport is inhibited, and another associated with the sudden onset of local instabilities that disrupt the wave regime and enhance energy dissipation \cite{Sujovolsky2019, Marino2022}. In this framework, Lagrangian particles that undergo extreme events correspond to those that temporarily escape the invariant manifold associated with internal waves, reaching regions of phase space where vertical transport and dissipation become more efficient. Conversely, particles that never experience extreme events remain confined within the wave-dominated manifold, undergoing more constrained evolution and reduced interaction with turbulent structures \cite{Sujovolsky2021}. This classification thus distinguishes between fluid elements that actively contribute to energy and mass mixing and those that persist in states of low mobility \cite{Sujovolsky2019}.

Based on these criteria, we define two subsets of particles: those that experience extreme events, denoted as $\#n_X$, and those that never experience them, denoted as $\#n_{NX}$. In each simulation, approximately $20\%$ of the total particles belong to $\#n_X$, while about $30\%$ belong to $\#n_{NX}$. These proportions yield two comparably sized groups, allowing for a robust statistical analysis of their contrasting dynamical behavior.

\section{Pair dispersion of tracers \label{sec:part_dist}}

The study of pair dispersion is central in turbulence, as it characterizes how flow structures influence the evolution of particle separations across scales \cite{Falkovich_2001, Salazar2009, Bourgoin2015}. In stratified flows, vertical dispersion is particularly important for atmospheric and oceanic dynamics, where stratification limits vertical transport and introduces anisotropy in particle separations \cite{Ollitrault2005, Sujo_lag_2019}.

Pair dispersion is analyzed through the time evolution of the relative separation $r(t)$ between two particles, governed by $dr/dt = w(t)$, where $w(t)$ is their relative velocity. The mean-square separation $\langle r^2 \rangle (t)$ characterizes different dispersion regimes depending on spatial and temporal scales. At small scales within the dissipative range, separation is controlled by local velocity gradients. In the inertial subrange ($\eta \ll r \ll L_0$), and assuming isotropic and homogeneous turbulence, dispersion follows the Richardson–Obukhov law, $\langle r^2 \rangle (t) \sim g \epsilon t^3$, where $\epsilon$ is the energy dissipation rate and $g$ a universal constant \cite{Bec2010, Bourgoin2015}. In stratified turbulence, particle dispersion is strongly anisotropic, with marked differences between horizontal and vertical components, including in the presence of rotation \cite{Gallon2025}. These effects lead to deviations from classical models developed for isotropic and homogeneous turbulence \cite{Salazar2009, Sujo_lag_2019}. Whereas horizontal dispersion displays sustained growth over time, vertical dispersion typically exhibits an initial growth phase, followed by saturation and eventually a transition to long-time diffusive behavior \cite{van_aartrijk_2008, Sujo_lag_2019}.

\begin{figure}
	\centering
	\includegraphics[width=0.46\textwidth]{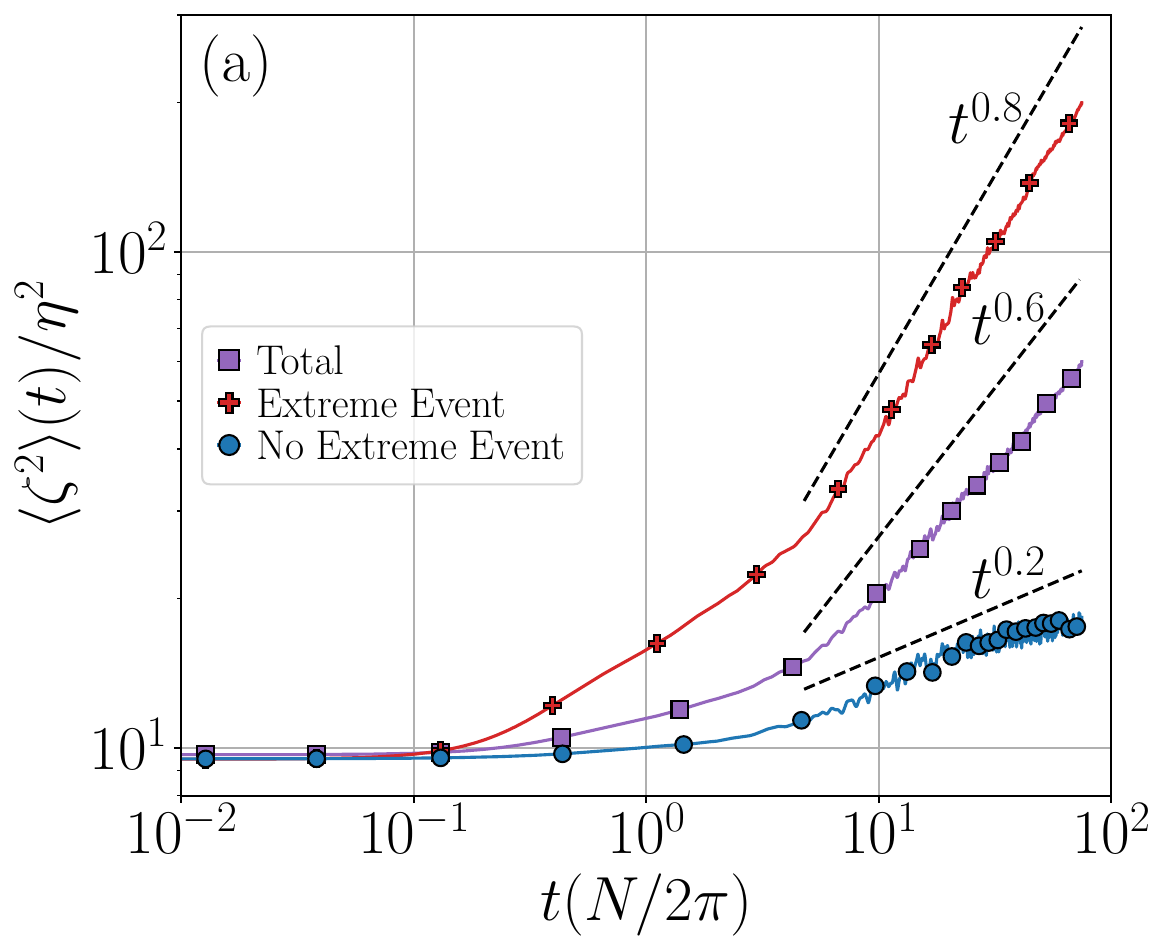}
	\includegraphics[width=0.46\textwidth]{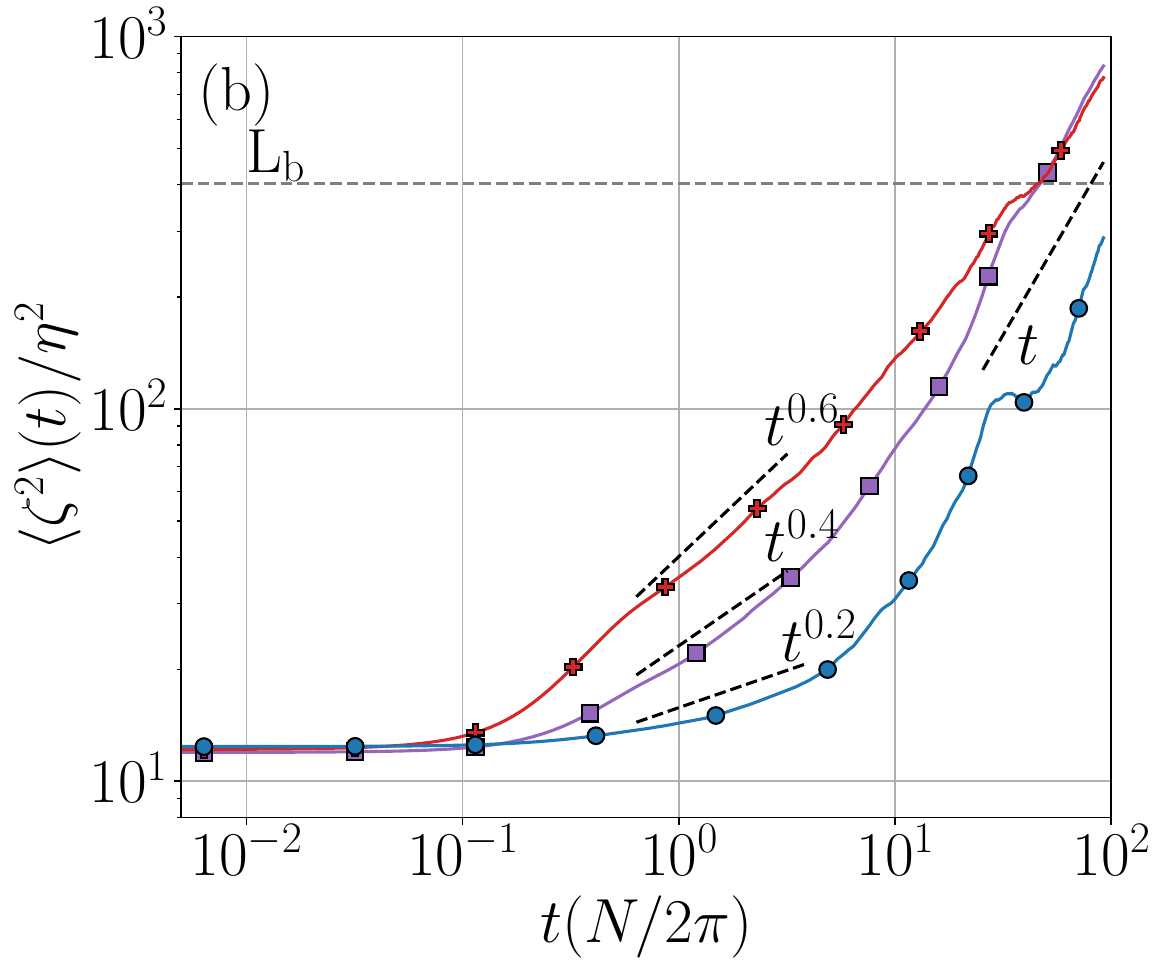}
	\caption{Vertical pair dispersion $\langle \zeta^2 \rangle (t)$, normalized by $\eta$, as a function of time for all tracers (purple), for the subset experiencing extreme events ($\#n_X$, red), and for the subset that does not ($\#n_{NX}$, blue). Panel (a) corresponds to the RND simulation, while panel (b) shows results for TG. In both cases, subset $\#n_X$ shows significantly enhanced dispersion, while $\#n_{NX}$ exhibits much lower values of $\langle \zeta^2 \rangle$. In the TG simulation, as the separation approaches the buoyancy scale, the subsets seem to transition to a diffusive regime.}
	\label{par_dis}
\end{figure}

Here we focus on vertical pair dispersion, considering its role in the vertical transport in geophysical turbulence. Figure \ref{par_dis} presents the temporal evolution of the vertical separation of tracer pairs initially separated by $\zeta_0 \sim 10\eta$, evaluated for the entire set of tracers as well as for its two subsets $\#n_X$ and $\#n_{NX}$ defined above. Dispersion is quantified through the mean-square vertical separation $\langle \zeta^2 \rangle (t)$, normalized by the dissipation scale $\eta$. Results from both RND and TG simulations are compared to examine how the forcing structure influences dispersion dynamics. In both cases, particles belonging to subset $\#n_X$---those undergoing extreme events---exhibit significantly larger dispersion than particles in subset $\#n_{NX}$, which never experience such events. This contrast is particularly evident in the dispersion exponent: although dispersion is diffusive in all cases (i.e., sublinear in time), the exponent is consistently larger for $\#n_X$. The qualitative difference in dispersion behavior is further illustrated in the supplementary video \cite{sm}, where two tracer pairs with identical initial separation evolve in the same flow: one pair encounters an extreme event and undergoes rapid vertical dispersion, while the other remains closely coupled. This contrast highlights the role of extreme events in driving conditional transport dynamics in stratified turbulence.

An interesting effect is observed in the TG simulation, where a change in the dispersion rate seems to occur as the separation approaches the buoyancy scale $L_b$. At that point, vertical dispersion transitions into a diffusive regime that is compatible with $\sim t$ scaling, suggesting a modification in the underlying transport dynamics. This behavior would be consistent with previous studies showing that stratification imposes an upper limit on vertical dispersion, regulating the interplay between turbulence and internal waves \cite{Sujo_lag_2019, Sujo_2018, van_aartrijk_2008}.

Moreover, the comparison between forcings reveals significant differences in the dispersion behavior. In the RND case, dispersion grows steadily in time, as expected since this forcing tries to generate a more isotropic turbulent flow without dominant structures that can restrict particle transport. In contrast, TG dispersion is strongly modulated by vertical shear, showing that flow organization directly influences vertical transport efficiency. In particular, the formation of large-scale coherent structures in TG tends to inhibit dispersion, confining particles and promoting accumulation in specific flow regions. These plateaus are associated with the layered structure characteristic of stratified flows and reflect the stratification-imposed constraints on vertical transport \cite{van_aartrijk_2008}. Overall, these differences highlight the fundamental role of flow structure in determining particle dispersion in stratified turbulence, and reinforce the importance of extreme events in mass mixing.

\begin{figure}
	\centering
	\includegraphics[width=0.46\textwidth]{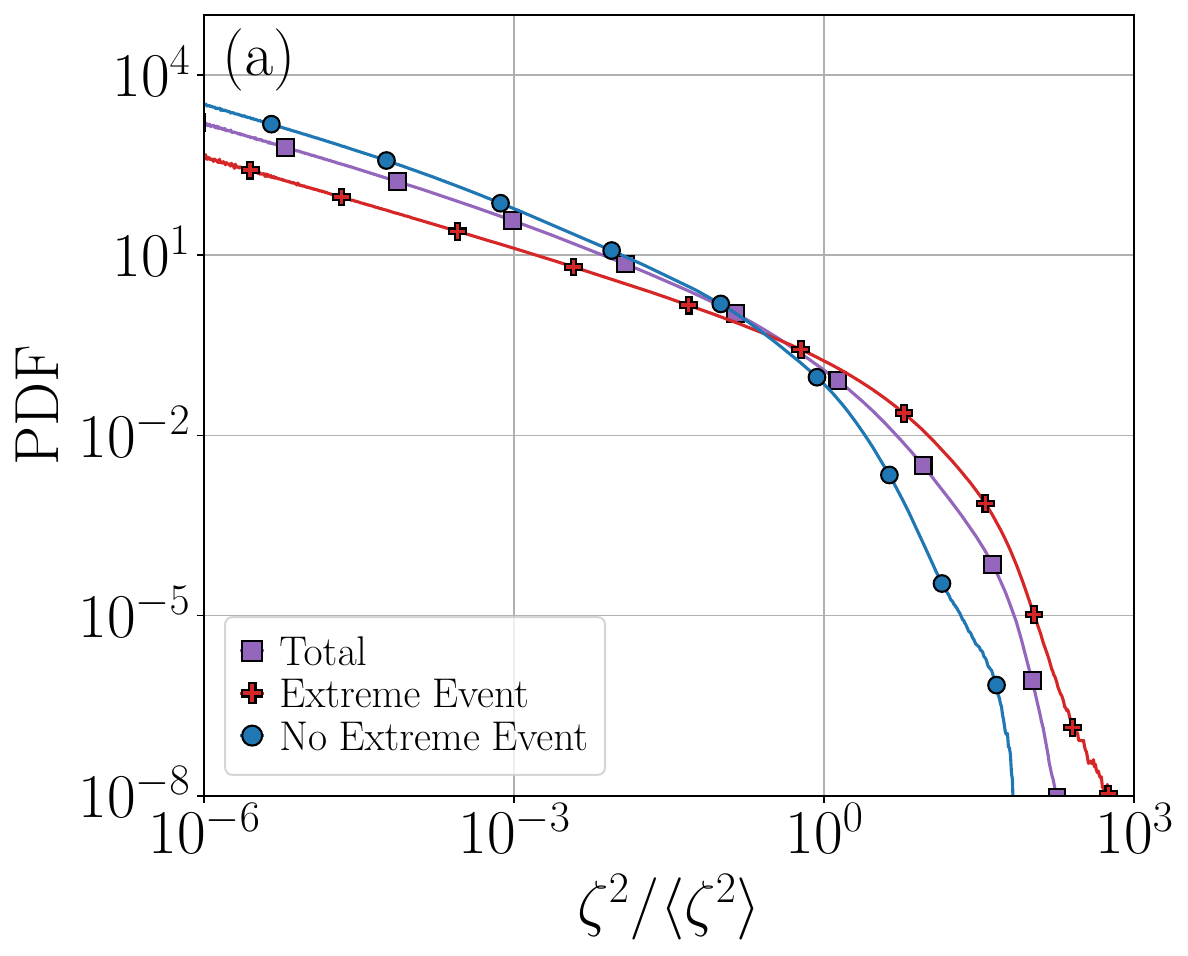}
	\includegraphics[width=0.46\textwidth]{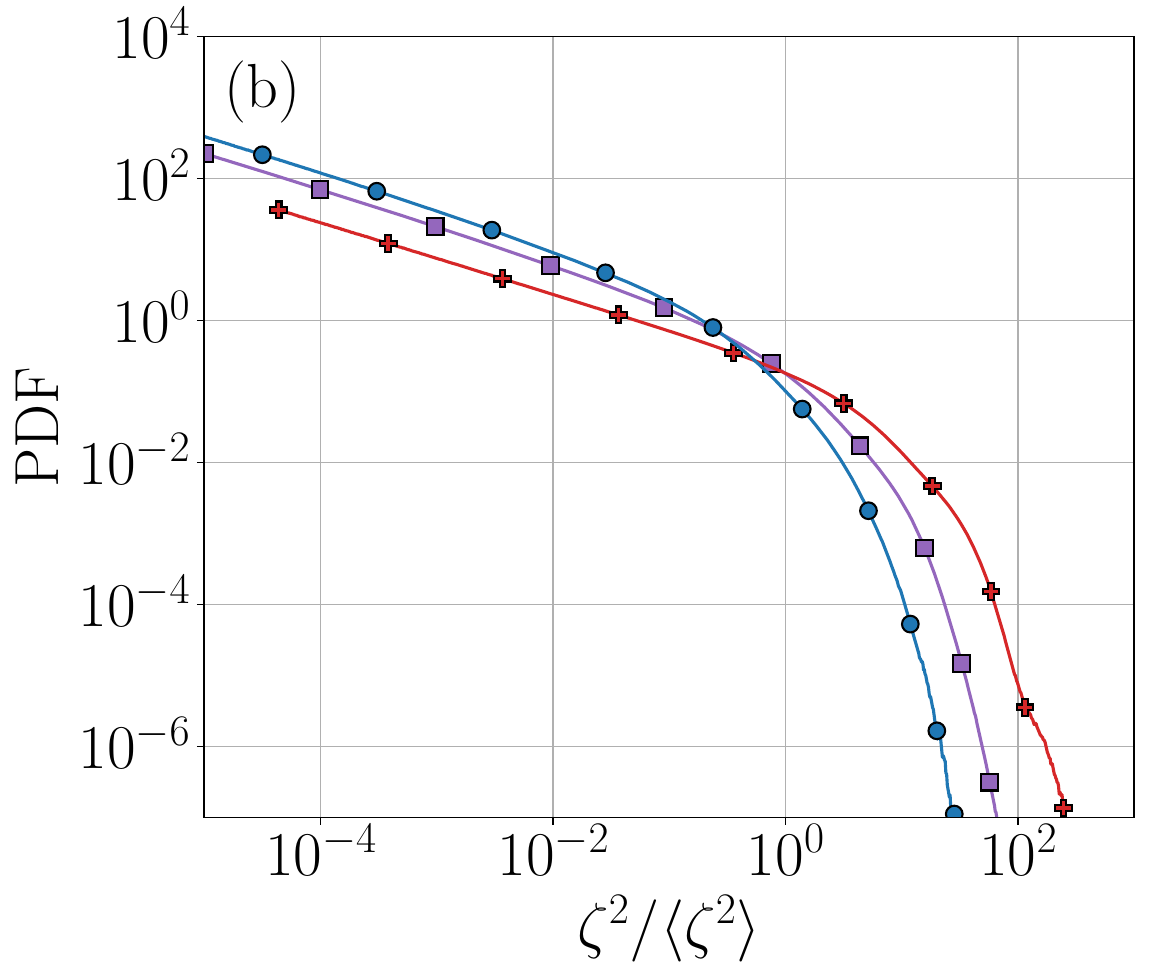}
	\caption{Probability density function (PDF) of vertical pair dispersion $\zeta^2$, normalized by its mean $\langle \zeta^2 \rangle$, for the full set of tracers, the subset that experiences extreme events ($\#n_X$), and the subset that does not ($\#n_{NX}$), evaluated at times larger than $\tau = 2\pi/N$. Panel (a) corresponds to the RND simulation, panel (b) to TG. In both cases, tracers in $\#n_X$ show larger probability of attain large separations, while tracers in $\#n_{NX}$ are more likely to remain below the mean.}
	\label{par_dis_pdf}
\end{figure}

To further examine pair dispersion in stratified turbulence, we analyze the distribution of vertical separations within each tracer subset. Figure~\ref{par_dis_pdf} shows the probability density function (PDF) of the vertical separation $\zeta^2$, normalized by its mean $\langle \zeta^2 \rangle$ at each time. Results are presented for the full tracer set, the $\#n_X$ subset, and the $\#n_{NX}$ subset, considering times larger than $\tau = 2\pi/N$. Figure~\ref{par_dis_pdf}(a) corresponds to RND, and Fig.~\ref{par_dis_pdf}(b) to TG.

In both cases, the PDFs exhibit extended tails, particularly for the $\#n_X$ subset. This reflects the presence of extreme separation events, where some tracers undergo vertical dispersion far exceeding the average. By contrast, the $\#n_{NX}$ subset displays a larger probability of separations below the mean, indicating that their vertical transport is more strongly constrained by stratification. In particular, the likelihood of finding pair separations below the mean is significantly larger for $\#n_{NX}$ than for $\#n_X$.

This behavior is similar to that observed in isotropic and homogeneous turbulence, where the PDF of pair separations departs from Gaussian statistics and exhibits heavy tails due to flow intermittency and the action of coherent structures \cite{Salazar2009, Rast_2016}. In stratified turbulence, vertical dispersion is further modulated by internal waves and vertical shear, introducing additional anisotropy in the statistics of tracer separations \cite{Sujo_2018, clark_di_leoni_2015}.

These results indicate that Lagrangian extreme events fundamentally reshape vertical pair dispersion in stratified turbulence. While classical theories such as the Richardson–Obukhov law predict universal scaling in isotropic and homogeneous turbulence, stratification imposes strong anisotropy and introduces intermittent events that drive substantial deviations from these models. The comparison between TG and RND further shows that the forcing structure regulates not only the mean level of dispersion but also the probability of extreme separations, with TG suppressing extremes through coherent shear layers and large-scale structures, and RND enabling their persistence in a slightly more isotropic setting. Overall, the Lagrangian classification into $\#n_X$ and $\#n_{NX}$ subsets provides a new perspective beyond global or Eulerian analyses, highlighting the essential role of extreme events in controlling vertical transport and mixing in stratified turbulence.

\section{Finite-time Lyapunov exponents and Lagrangian coherent structures}

As a result of the previous discussion, it seems evident that understanding particle transport in stratified turbulent flows requires an accurate characterization of the structures that organize the dynamics. Eulerian methods can identify vortices and structures in the instantaneous velocity field but are limited when analyzing transport evolution over time. In this context, finite-time Lyapunov exponents (FTLEs) and Lagrangian coherent structures (LCSs) provide a rigorous mathematical framework for identifying material surfaces that act either as transport barriers or as regions of enhanced mixing \cite{haller2015}.

The study of LCSs and FTLEs has advanced considerably over the past decades, providing essential tools for analyzing transport dynamics in turbulent flows. Early efforts to characterize such structures in chaotic and turbulent systems trace back to Aref \cite{Aref1984}, who showed how chaos theory applied to fluids can be used to understand mixing and the organization of transport in time-dependent flows. Subsequent work by Ottino and Pierrehumbert extended these ideas to chemical and geophysical flows, demonstrating that coherent structures in phase space strongly influence tracer dispersion \cite{Pierrehumbert1991, Ottino1989}.

The modern formulation of LCSs, based on the maximization of FTLEs, was introduced by Haller \cite{Haller2001, Haller2002}, establishing a mathematical framework to identify these structures in unsteady flows. These studies showed that LCSs can be classified into attracting and repelling structures depending on the local stretching dynamics of the fluid. This approach has been extensively validated through both numerical and experimental studies, demonstrating that regions of high FTLE typically coincide with transport barriers in turbulent systems \cite{Shadden2005}. In the context of stratified turbulence, works by Boffetta et al.~and Ruppert-Felsot et al.~analyzed how internal waves and vertical shear modify the structure of LCSs \cite{Boffetta2001, Mathur2007}. Their results showed that stratification introduces anisotropy in transport dynamics, with coherent structures in stratified flows displaying greater temporal persistence than in isotropic turbulence. More recently, Haller and Beron-Vera applied this framework to oceanic vortices, identifying LCSs as dynamic barriers that regulate large-scale mass dispersion \cite{Haller2012}.

The approach to detect LCSs is then based on the computation of FTLEs, which quantify the exponential rate of separation between initially nearby trajectories in the flow. Mathematically, the FTLE is defined from the separation of particle pairs as
\begin{equation}
\lambda(\zeta_0, t_0, t) = \frac{1}{|t - t_0|} \ln{\left(\frac{\zeta(t)}{\zeta_0}\right)},
\end{equation}
where $\lambda$ is the Lyapunov exponent. Large FTLE values identify regions where small perturbations in the initial tracer or inertial particle positions lead to large trajectory divergences, indicating the presence of attracting or repelling LCSs \cite{haller2015}.

\begin{figure}
	\centering
	\includegraphics[width=0.46\textwidth]{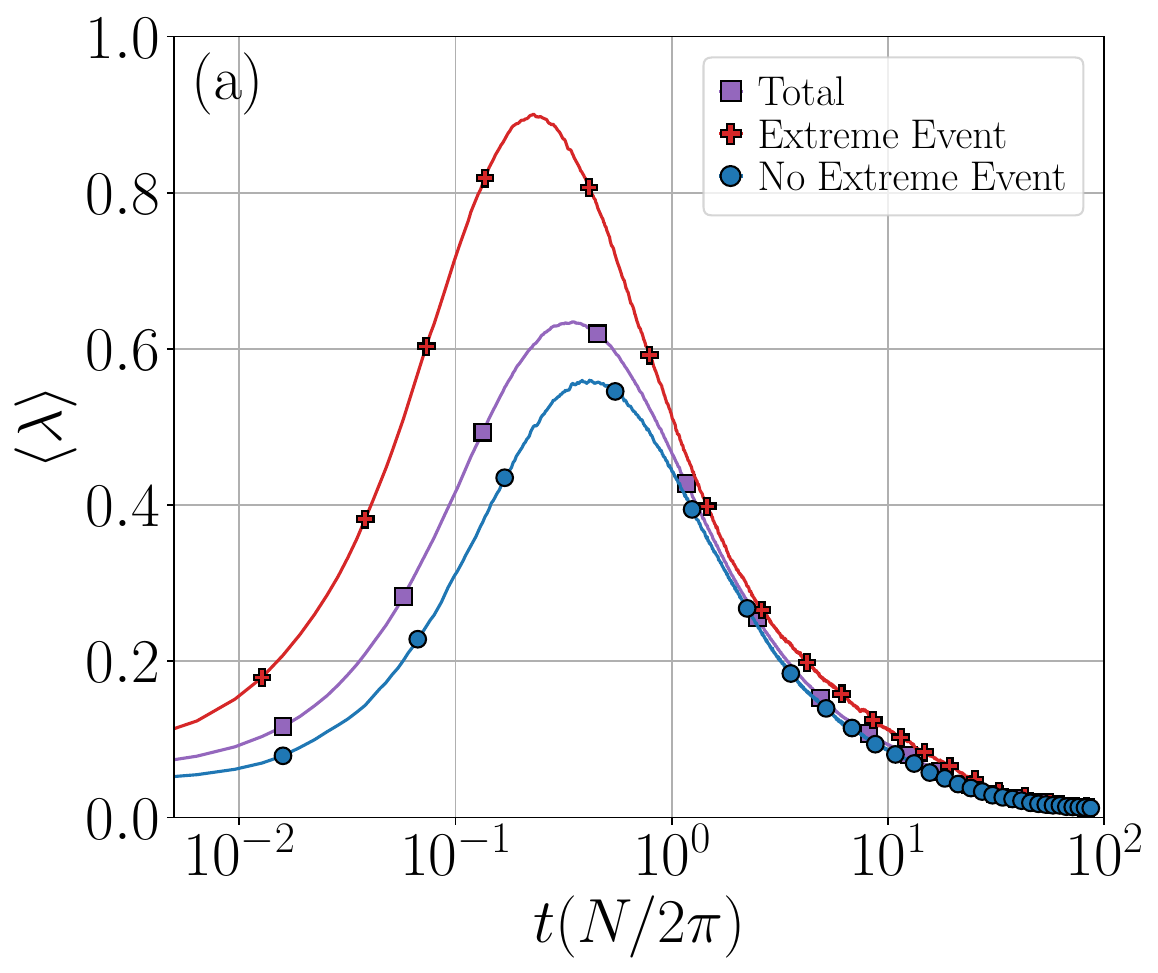}
	\includegraphics[width=0.46\textwidth]{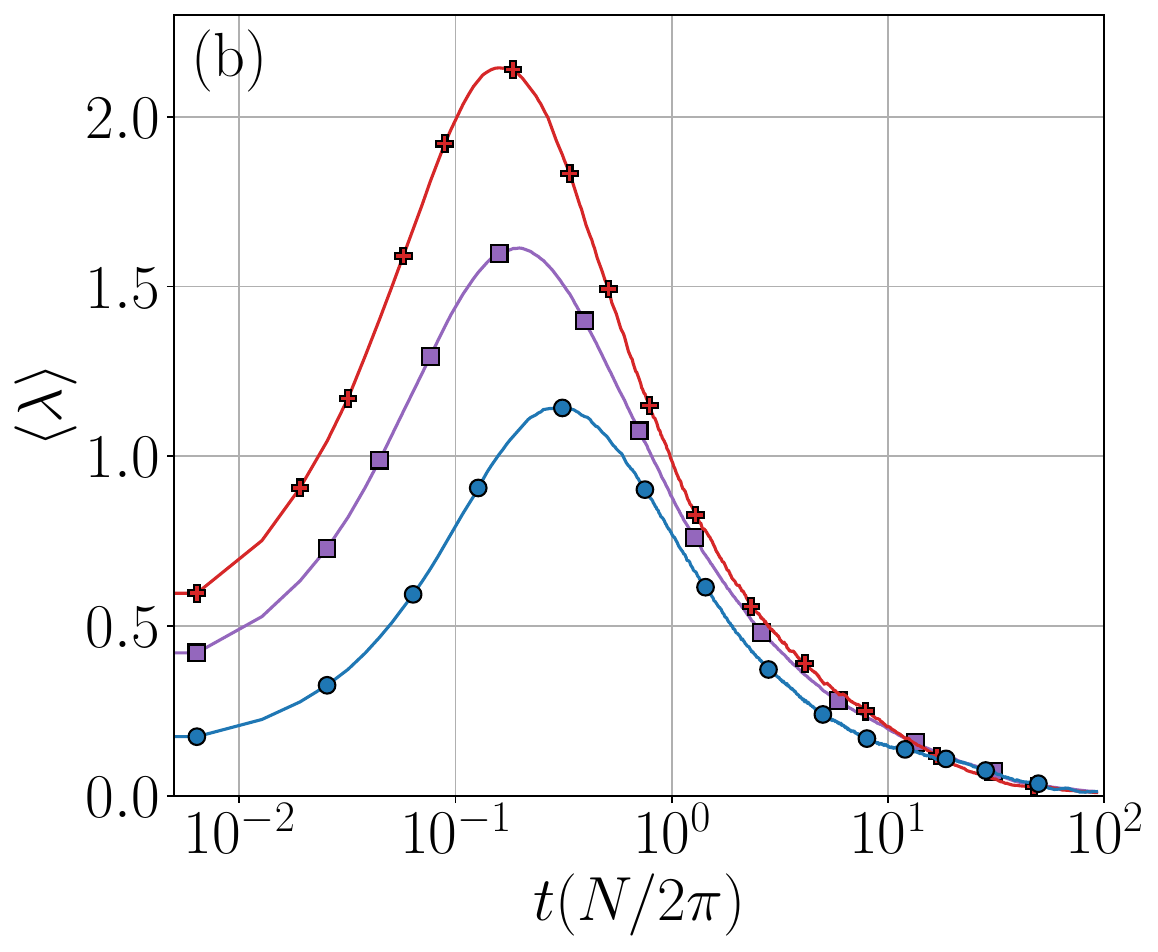}
	\caption{Mean finite-time Lyapunov exponent, $\langle \lambda \rangle$, normalized by the Brunt–Väisälä frequency, as a function of time for all tracers, the subset that experiences extreme events ($\#n_X$), and the subset that does not ($\#n_{NX}$). Panel (a) corresponds to the RND simulation, while panel (b) shows results for the TG simulation. In both cases, $\#n_X$ shows significantly larger $\langle \lambda \rangle$, indicating a larger prevalence of repelling trajectories within this subset.}
	\label{lyapunov}
\end{figure}

Figure~\ref{lyapunov} shows the temporal evolution of the mean FTLE for the full tracer set, as well as for the subsets $\#n_X$ and $\#n_{NX}$. Panel (a) corresponds to the RND simulation, while panel (b) presents the results for TG. In both cases, $\#n_X$ exhibits significantly larger $\langle \lambda \rangle$ values, indicating a stronger prevalence of repelling trajectories within this group. At long times, $\langle \lambda \rangle$ tends toward zero, as tracer pairs reach large separations and no longer undergo exponential divergence. At short times, a maximum in $\lambda$ appears near the ballistic separation scale, corresponding to the regime where tracer motion is still dominated by initial velocities, leading to rapid, nearly linear separation \cite{Sujo_2018, Reartes2023}.

The fact that subset $\#n_X$ attains the largest $\lambda$ values reveals a clear correlation between extreme events and strongly repelling trajectories. This effect is even more pronounced in the TG simulation, where the Lyapunov exponent reaches values nearly twice those observed in the RND case. This difference suggests that, in the TG flow, tracer dynamics are dominated by coherent structures that intensify trajectory divergence and amplify dispersion.

\begin{figure}
	\centering
	\includegraphics[width=0.46\textwidth]{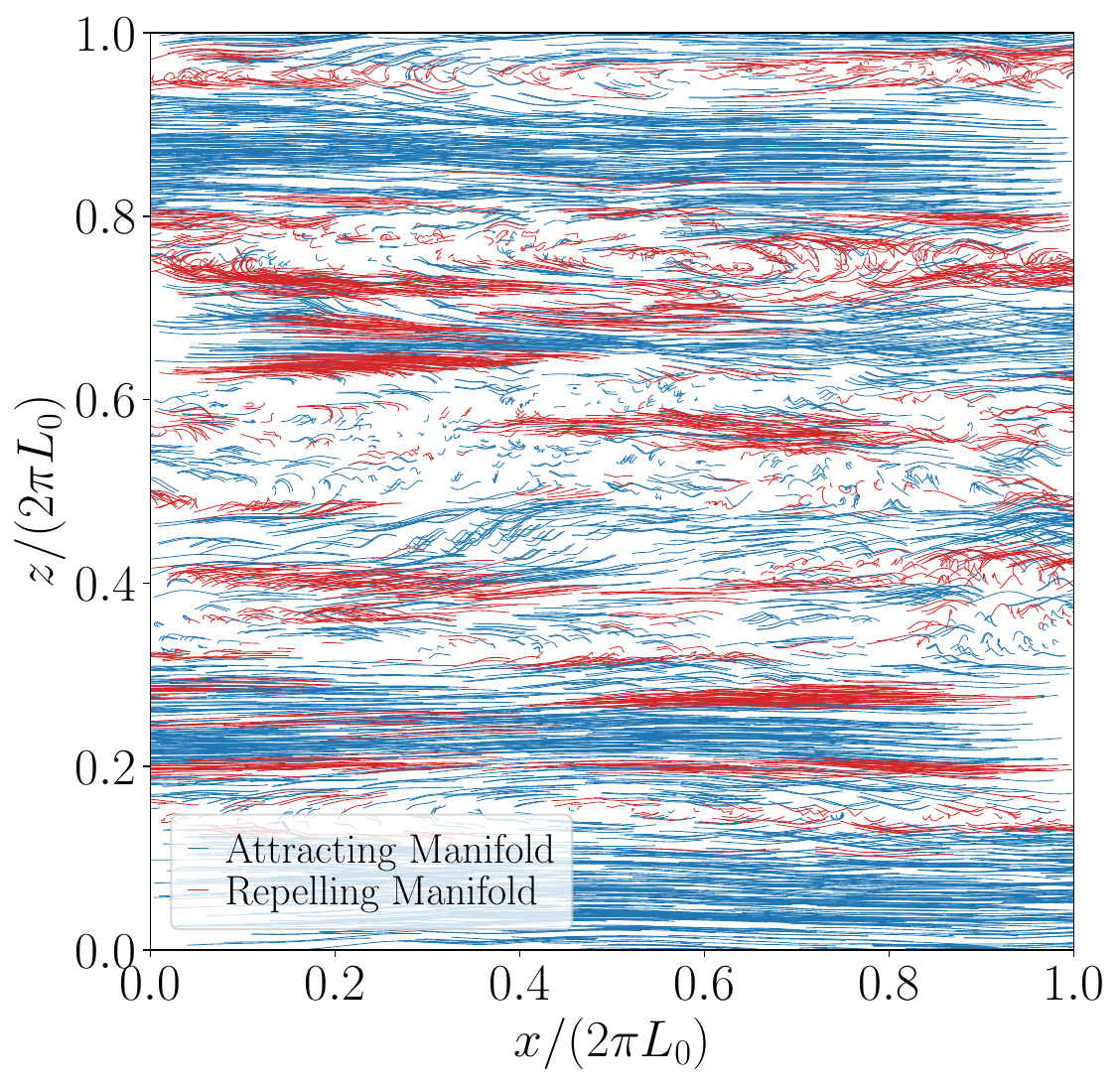}
	\includegraphics[width=0.46\textwidth]{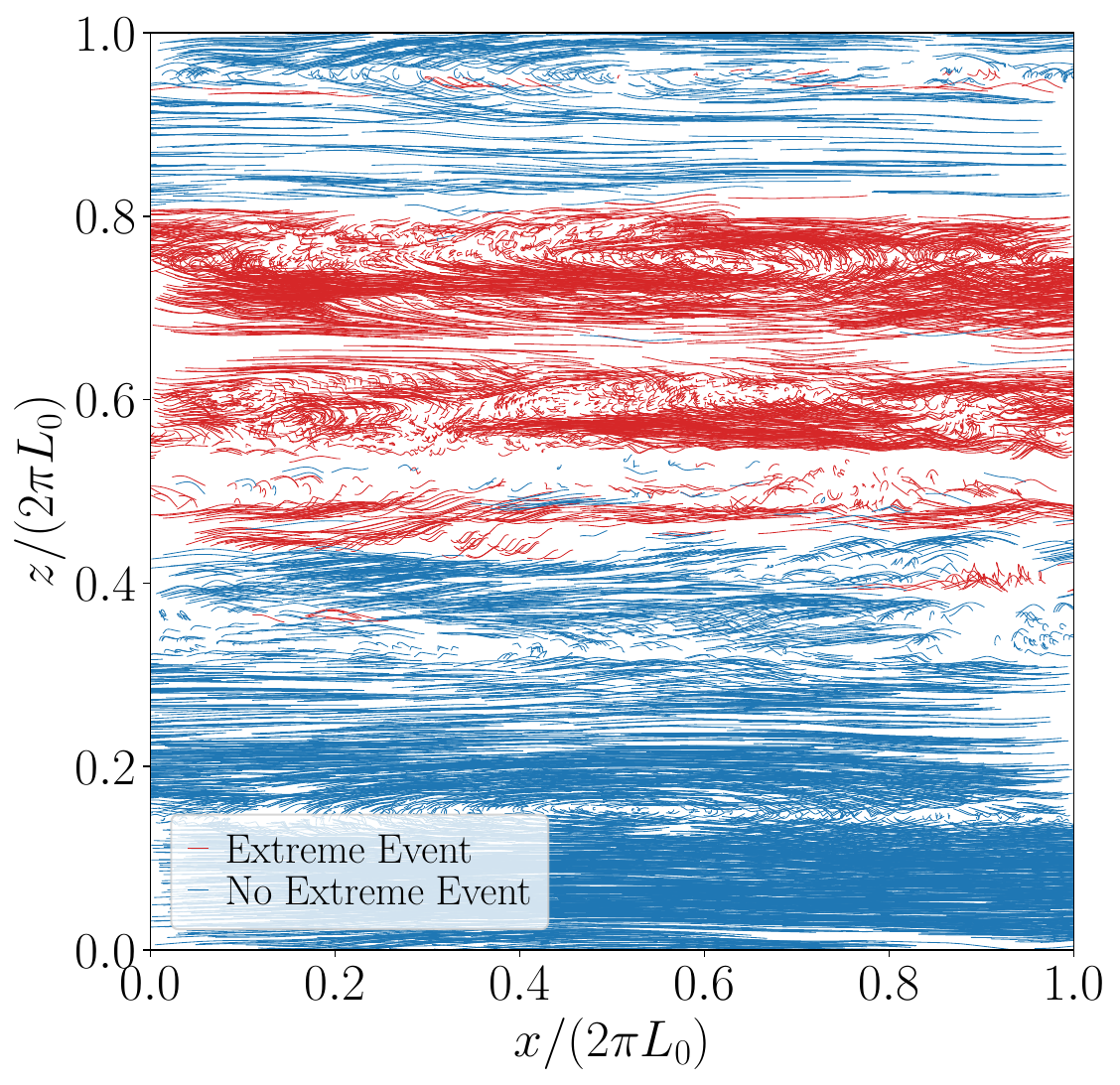}
	\caption{Tracer trajectories over a time interval $\tau = 2\pi/N$ for the RND simulation. Panel (a): trajectories colored by FTLE values---red indicates repelling structures (positive FTLE), blue indicates attracting structures (negative FTLE). Panel (b): same trajectories colored by whether tracers experienced extreme events (red) or not (blue). A strong correlation is observed between repelling trajectories and the occurrence of extreme events.}
	\label{map_1024}
\end{figure}

\begin{figure}
	\centering
	\includegraphics[width=0.65\textwidth]{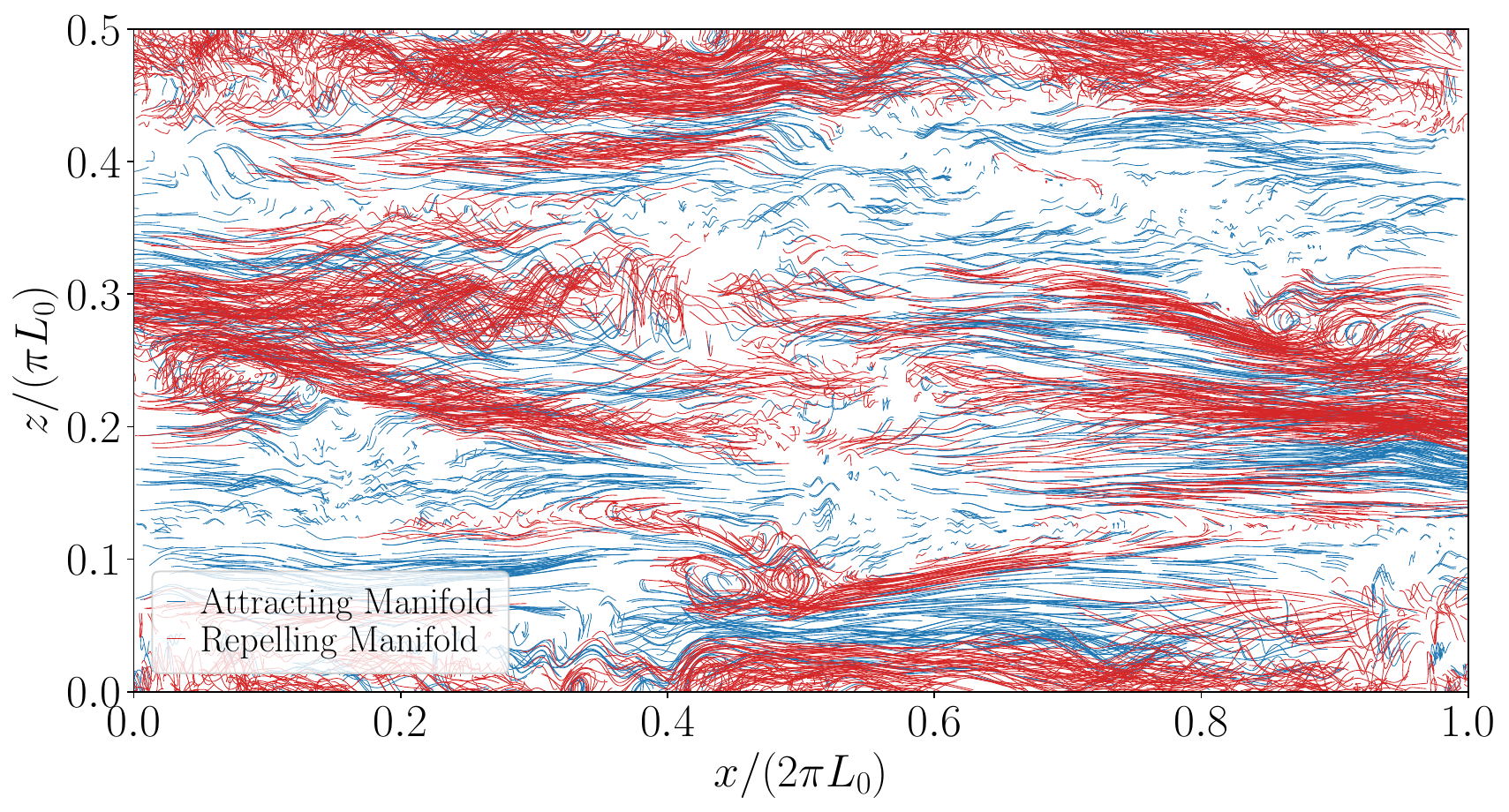}
	\includegraphics[width=0.65\textwidth]{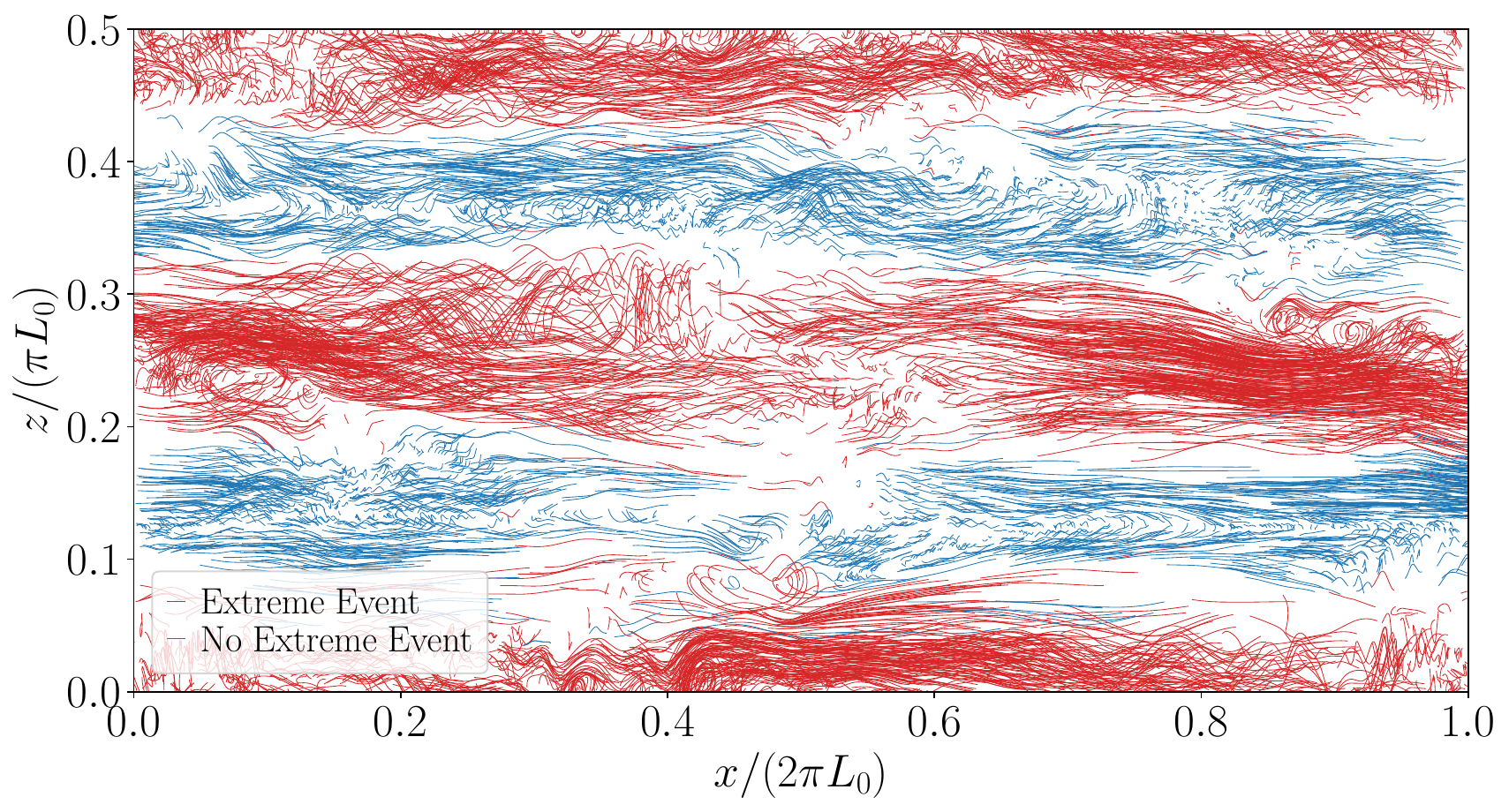}
	\caption{Tracer trajectories over a time interval $\tau = 2\pi/N$ for the TG simulation. Panel (a): trajectories colored by FTLE values---red indicates repelling structures (positive FTLE), blue indicates attracting structures (negative FTLE). Panel (b): same trajectories colored by whether tracers experienced extreme events (red) or not (blue). An even stronger correlation is observed compared to the RND case, suggesting that repelling trajectories are closely associated with extreme events.}
	\label{map_768}
\end{figure}

Figures~\ref{map_1024} and \ref{map_768} show tracer trajectories in the RND and TG simulations on an $x$–$z$ slice, highlighting the relationship between Lagrangian coherent structures and the occurrence of extreme events. In each case, panel (a) colours trajectories according to FTLE values---distinguishing repelling (positive FTLE) and attracting (negative FTLE) regions---while panel (b) classifies trajectories by whether tracers experienced extreme events. A clear spatial correlation emerges between repelling trajectories and those associated with extreme events, suggesting that LCSs play a central role in modulating vertical transport in these flows. This is consistent with previous studies showing that regions of high FTLE coincide with zones of intense flow stretching, where small initial differences in tracer positions grow exponentially, enhancing trajectory separation \cite{haller2015}. However, the results presented here also correlate these structures with the occurrence or not of extreme events.

In the TG simulation, the presence of coherent structures is even more evident, with well-defined repelling regions closely correlating with high concentrations of extreme events. This reinforces the view that LCSs not only organize flow transport but also act as catalysts for the occurrence of extreme events in particle dispersion. The clustering of extreme vertical drafts in specific regions of the flow further suggests the existence of high-intermittency patches, akin to those observed in atmospheric turbulence \cite{mahrt_89,rodriguezimazio2023}. These patches, characterized by elevated dissipation rates and intense velocity fluctuations, may locally enhance dispersion, strengthening the link between coherent structures and intermittent transport.

In stratified turbulence, these structures can act either as dynamic barriers that inhibit mixing in certain regions or as preferential corridors that promote material exchange between fluid layers \cite{haller2015}. The observation that repelling trajectories identified by FTLEs are associated with extreme events indicates that these regions exert a significant influence on vertical dispersion dynamics. In the TG simulation---where vertical shear and organized structures dominate the dynamics \cite{Sujovolsky2018}---this effect is more pronounced, which may explain the greater prevalence of highly unstable trajectories compared with the RND case. These findings strengthen the connection between extreme events and coherent flow structures, providing a Lagrangian framework to interpret their role in turbulent mixing.

\section{Relationship between extreme events and energy dissipation}

The study of energy dissipation in stratified turbulence has shown that the dynamics of these flows are dominated by highly intermittent regions, where both the dissipation rate and the enstrophy attain values far exceeding those in the rest of the domain. These regions of strong large-scale intermittency, corresponding to extreme drafts, have been identified in numerical simulations and atmospheric observations, and play a key role in redistributing energy and in regenerating turbulence in these systems \cite{Pearson_2018, Smyth_2019, Marino2022, rodriguezimazio2023, Foldes2025}.
In particular, recent studies have shown that the occurrence of extreme vertical bursts is a fundamental mechanism behind the formation of these regions. These bursts, which appear intermittently in space and time, can trigger localized turbulence, enhancing energy dissipation and shaping its statistical distribution \cite{Marino2022}. In geophysically relevant regimes, it has been observed that roughly $10\%$ of the flow volume can account for up to $50\%$ of the total dissipation, consistent with observations in ocean models \cite{Pearson_2018}, and underscoring the importance of these events in the energy budget of stratified turbulence \cite{Feraco_2018}.

In stratified flows, enstrophy can be also amplified in regions where shear and stratification produce thin turbulent layers \cite{Feraco_2018}. These layers act as zones of intense mixing and energy transfer, favouring the emergence of localized dissipation events. Numerical studies have shown that the link between enstrophy and extreme intermittency is even more pronounced in flows with well-defined coherent structures. In particular, in the TG simulation, particles undergoing extreme events exhibit a significant increase in enstrophy, reaching values far higher than in the RND case. This behavior suggests that the coherent structures characteristic of TG forcing not only organize flow dynamics but also amplify local intermittency and energy dissipation. In this sense, the presence of coherent vortices may facilitate the formation of high-shear regions, promoting the emergence of intermittent patches.

From an experimental perspective, the existence of high-intermittency patches has been confirmed through atmospheric turbulence measurements. Data collected aboard the HALO research aircraft show that clear-air turbulence exhibits a highly fragmented character, with well-defined regions of intense dissipation separated by zones of lower activity \cite{rodriguezimazio2023}. These patches display different spectral signatures within the same event, indicating that multiple mechanisms modulate dissipation across distinct regions. In particular, the correlation between vertical velocity and potential temperature has been observed to play a central role in the generation of these structures, suggesting that the interaction between internal waves and turbulence is a fundamental mechanism behind their onset.

\begin{figure}
	\centering
	\includegraphics[width=0.46\textwidth]{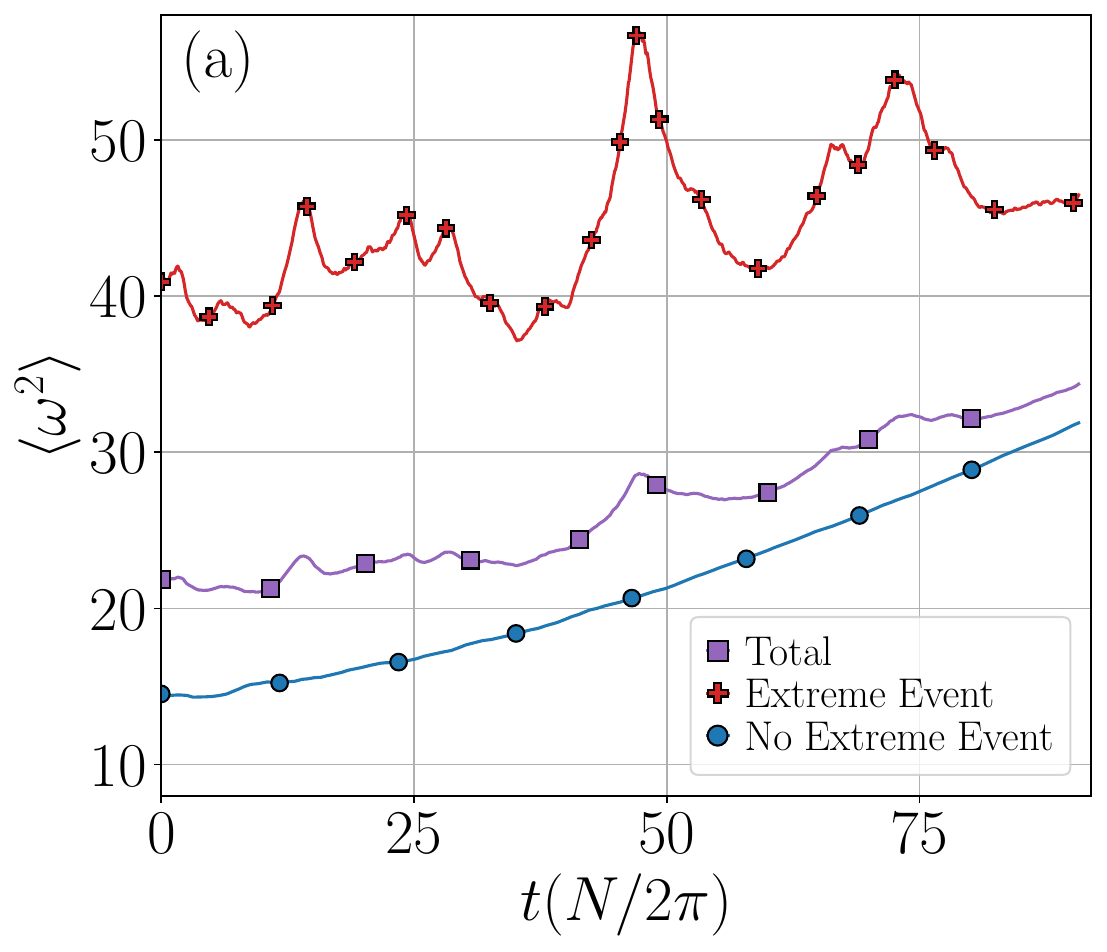}
	\includegraphics[width=0.471\textwidth]{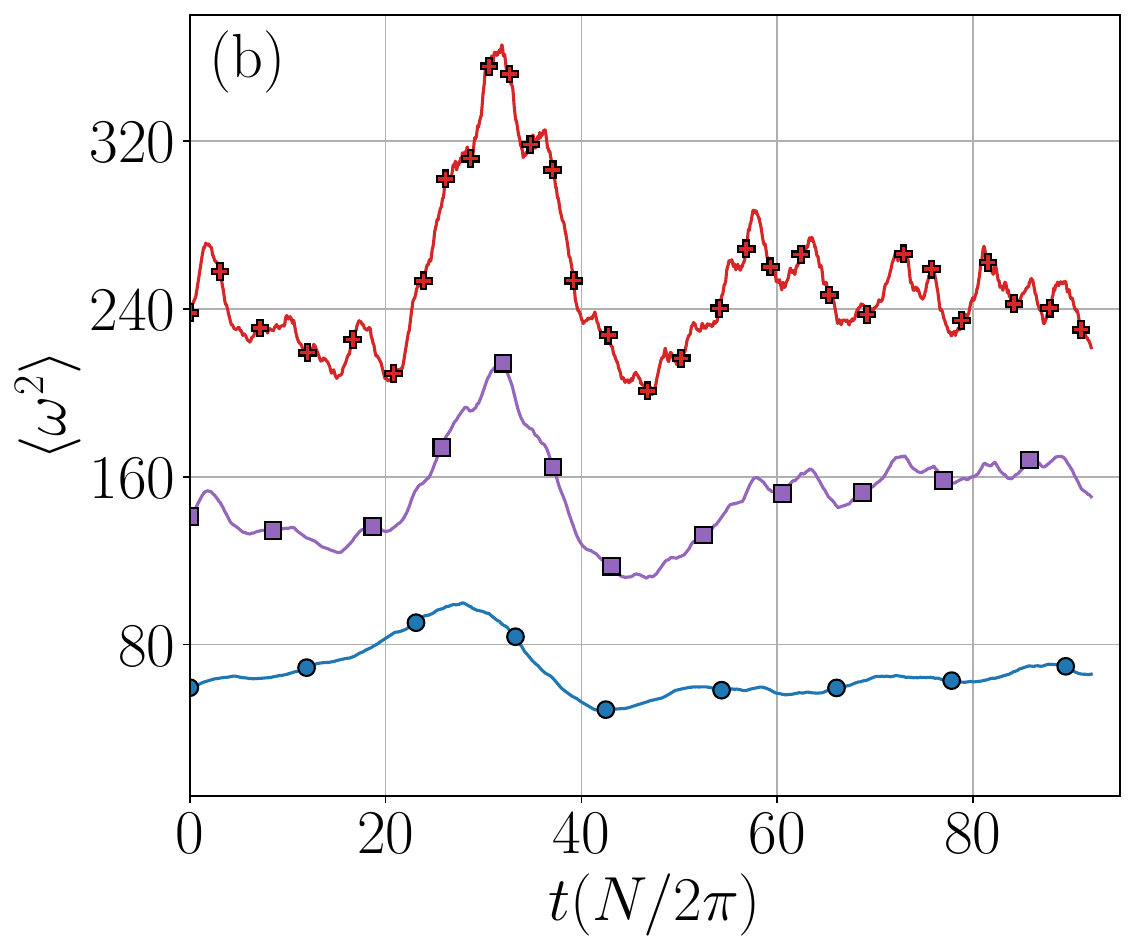}
	\caption{Temporal evolution of the mean squared vorticity at the positions of Lagrangian tracers for the full set, the subset that experiences extreme events ($\#n_X$), and the subset that does not ($\#n_{NX}$). Panel (a) corresponds to the RND simulation, while panel (b) shows the results for the TG simulation. In both cases, $\#n_X$ displays significantly larger squared vorticity than the average, while $\#n_{NX}$ shows markedly lower values. This suggests that extreme events predominantly occur in flow regions characterized by enhanced turbulence intensity.}
	\label{ens_exev}
\end{figure}

We also examined the link between enstrophy, energy dissipation, and extreme events in the simulations of stratified flows performed for this study, and vertical bursts were found to significantly affect the efficiency of kinetic energy dissipation. The spatial peaks of squared vorticity confirm that the extreme vertical drafts not only generate local turbulence, but also modulate energy dissipation and interscale transfer as found in \cite{Feraco_2018, Marino2022, Foldes2025}.
Figure \ref{ens_exev} presents the temporal evolution of the mean squared vorticity at the locations of Lagrangian tracers, for the total set as well as for the subsets that do or do not experience extreme events. The subset of particles undergoing extreme events exhibits significantly larger squared vorticity values, indicating that turbulence is stronger and energy dissipation is probably highly localized in these regions of the flow. This effect is even more pronounced in the TG simulation, where coherent structures organize the flow such that squared vorticity peaks coincide with highly intermittent regions. These results reinforce the view that large-scale intermittent turbulent patches are a fundamental feature of stratified turbulence, and play a central role in redistributing energy across the system \cite{Marino2022}. By concentrating the vorticity in localized regions, these patches strongly influence tracer transport dynamics and the evolution of turbulence at all scales.

It is also worth mentioning that in Fig.~\ref{ens_exev} there is another conspicuous difference between the RND and TG runs. In the TG flow, $\langle \omega^2 \rangle$ fluctuates around a mean value, while in RND it grows in time. This is associated to the dynamics of VSHW: while in the TG flow horizontal winds are constrained to the central shear layer, in the RND flow they can grow everywhere. The RND flow also has more energy in waves, that can feed the VSHW \cite{Smith_2002b}. As a result, this flow never reaches a truly stationary state, with more and more energy piling up in the VSHW modes. This results in the growth of $\langle \omega^2 \rangle$ seen in Fig.~\ref{ens_exev}(a).

To quantify how extreme events bias the overall dissipation budget, we tracked the instantaneous dissipation sampled by each tracer along its trajectory and integrated it over the full observation window for each subset. We find that the subset $\#n_{\mathrm X}$---which comprises only ${\sim}20\%$ of all tracers and, consequently, a comparable fraction of the sampled flow volume---accounts for ${\sim}40\%$ of the cumulative viscous dissipation. Put differently, tracers traversing extreme vertical drafts experience, on average, a dissipation rate nearly twice that of the background population. This disproportionate contribution confirms that extreme-event patches not only localize intense squared vorticity but also channel a substantial fraction of the energy cascade into small-scale dissipation. Unlike previous Eulerian analyses, this Lagrangian perspective reveals how a minority of fluid elements concentrate a dominant share of the dissipation. Thus, extreme events exert an outsized influence not only on transport and mixing efficiency, but also on the global energy budget.

\section{Effect of extreme events on the dispersion of inertial particles}

Here we briefly analyze how coherent structures influence the vertical stability of neutrally buoyant inertial particles. These particles have the same density as the surrounding fluid and therefore, in an isotropic and homogeneous fluid, experience no net buoyant force \cite{Bec2009}. However, in a stratified flow, as soon as they move from the layer of neutral buoyancy, they feel a restoring force. Besides, owing to their finite inertia, they do not behave as passive tracers, as they cannot instantly follow fluid accelerations \cite{van_aartrijk_2008, Bec2010}. In stratified flows, such particles tend to remain confined within a vertical region of thickness set by the buoyancy scale $L_b$, oscillating around isopycnal surfaces that match their density. The stability of this layer enables mostly horizontal transport, driven by vertically sheared horizontal winds \cite{Reartes2023}. The present analysis investigates instead the impact of extreme vertical velocity events on the stability of these particles within isopycnal layers, by examining their vertical dispersion. Since neutrally buoyant inertial particles generally remain near regions of matching density, it is unlikely that they encounter extreme events randomly. As discussed earlier, such events are not uniformly distributed throughout the flow but often extend across planes in the direction perpendicular to gravity \cite{feraco2021connecting}. To ensure exposure to regions of high extreme-event activity, we thus consider only the TG flow, and initialize particles in the center of the domain, where the Taylor–Green forcing generates strong vertical mixing and extreme events are highly concentrated (see Fig.~\ref{map_768}).

With this setup, we find that only $5\%$ of the particle sets listed in Table~\ref{tab_parts} do not experience any extreme events ($\#n_{NX}$). We define the subset $\#n_X$ as the group of particles that undergo extreme events, using the same strict criterion as for Lagrangian tracers. In this case, this subset comprises approximately $15\%$ of all particles. This lower fraction is expected, since inertia causes the particles to filter fast fluctuations in the fluid velocity and to deviate from the fluid motion: upon crossing an extreme event, inertial particles tend to remain trapped in the flow for a shorter time, and stratification-induced confinement leads them to decouple from the fluid element more rapidly.

\begin{figure}
\centering
\includegraphics[width=0.48\textwidth]{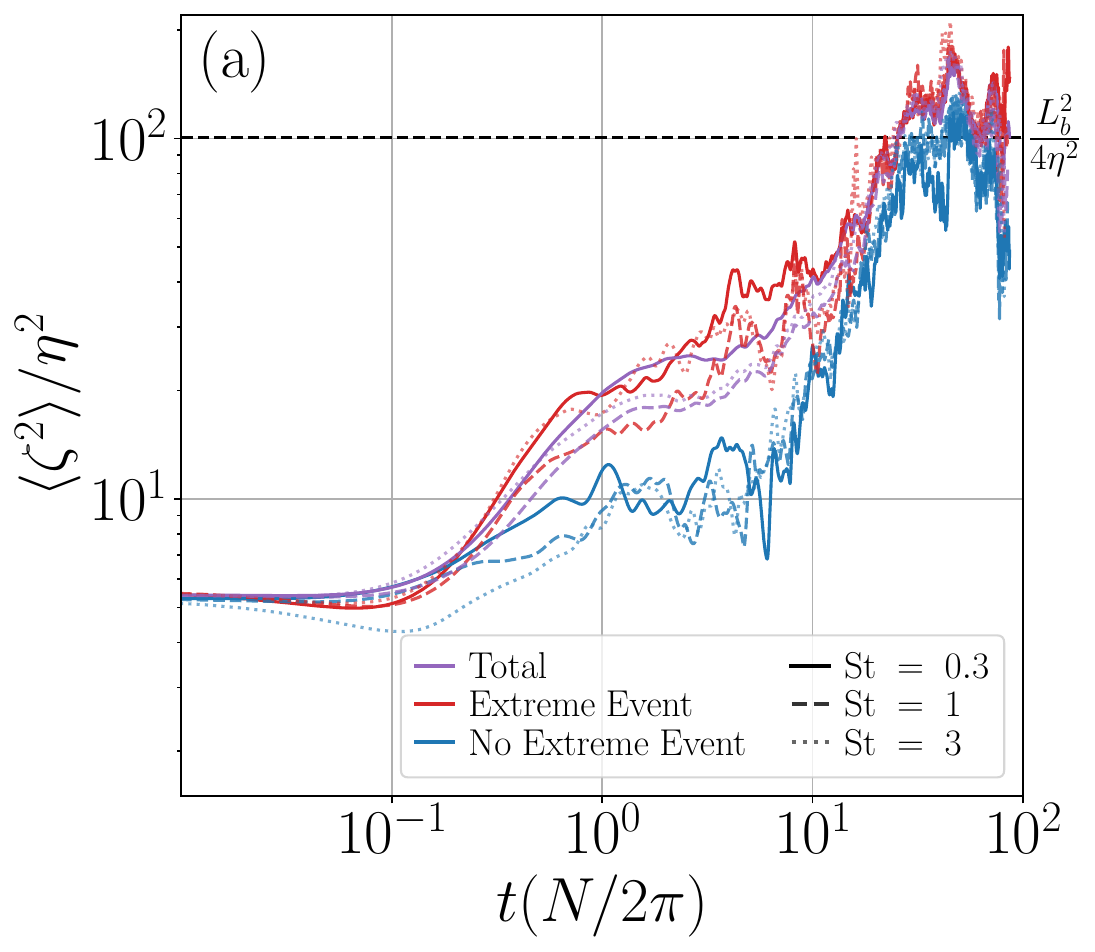}
\includegraphics[width=0.405\textwidth]{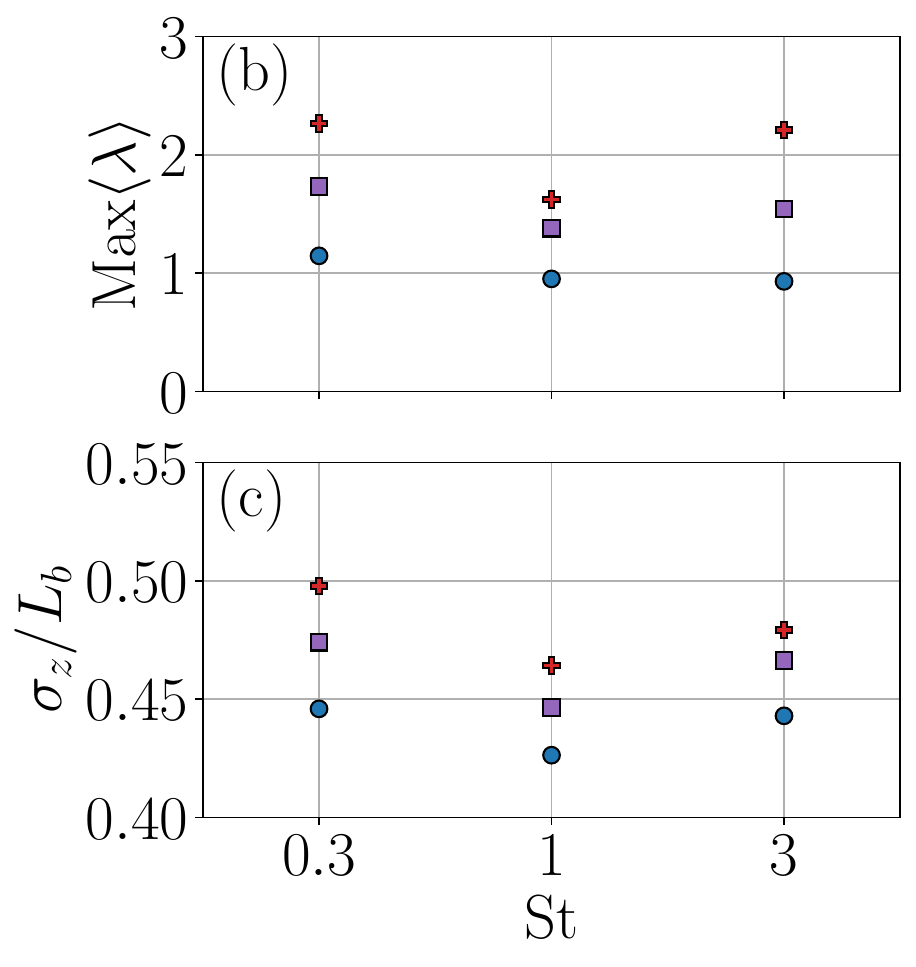}
\caption{(a) Vertical pair dispersion $\langle \zeta^2 \rangle (t)$, normalized by $\eta^2$, as a function of time for different Stokes numbers ($\textrm{St}$) of inertial particles. Extreme events have a significant impact on dispersion, particularly for particles with low inertia. At long times, the pair dispersion saturates at a value proportional to the buoyancy scale $L_b$. At short times, particles undergoing extreme events exhibit a rapid initial separation. (b) The finite-time Lyapunov exponent, normalized by the Brunt–Väisälä frequency, reflects this fast separation, showing markedly larger values for particles experiencing extreme events, especially for lower Stokes numbers. (c) The standard deviation of particle vertical positions, $\sigma_z$, normalized by $L_b$, highlights the crucial role of extreme events in vertical dispersion. This influence is stronger for particles with lower inertia, and becomes less significant as $\textrm{St}$ increases.}
\label{iner}
\end{figure}

Figure \ref{iner}(a) shows the vertical pair dispersion $\langle \zeta^2 \rangle (t)$ of inertial particles, initially separated by $\zeta_0 \sim 5\eta$, normalized by $\eta^2$, as a function of time for different Stokes numbers. Dispersion for particles in the $\#n_X$ subset is significantly larger than the average, whereas the $\#n_{NX}$ subset exhibits much lower values. For $\textrm{St} = 0.3$, dispersion is particularly enhanced compared with higher-inertia cases, as particles with small $\textrm{St}$ are more strongly coupled to the fluid elements. At long times, all curves converge to a value proportional to $L_b/2$, corresponding to the mean separation of a randomly distributed particle pair within the stratified layer of thickness $L_b$. This behavior is consistent with the vertical confinement imposed by stratification. Nevertheless, even at long times the $\#n_X$ subset exhibits larger dispersion than $\#n_{NX}$, indicating that particles undergoing extreme events have a larger tendency to explore the flow outside the layer where they are otherwise stably confined. Moreover, the dispersion curve for $\#n_X$ closely matches the ensemble average, consistent with the observation that in this configuration most particles undergo at least one extreme event during their evolution, as discussed previously.

To assess the short-time separation rate of inertial particle pairs, Figure~\ref{iner}(b) presents the maximum finite-time Lyapunov exponent for different particle subsets and Stokes numbers ($\textrm{St}$). In all cases, the $\#n_X$ subset---particles experiencing extreme events---consists of highly repelling trajectories, whereas the $\#n_{NX}$ subset shows significantly fewer. This effect is particularly pronounced for particles with $\textrm{St} = 1$, a value known to correspond to maximum particle clustering \cite{Aartrijk_prefential_2008,Reartes2023}, which occurs in regions of low local vorticity due to strong particle–fluid coupling. In this case, the difference between the Lyapunov exponent of the $\#n_X$ subset and the global average is markedly smaller, suggesting that most particles at $\textrm{St} = 1$ already follow trajectories with enhanced separation dynamics.

Another remarkable result for neutrally buoyant inertial particles concerns their stability within the buoyancy layer. Figure~\ref{iner}(c) shows the standard deviation of particle vertical positions, $\sigma_z$, normalized by $L_b$. Particles in the $\#n_X$ subset exhibit significantly larger dispersion than the mean, while those in $\#n_{NX}$ display much smaller values. This indicates that extreme events play a destabilizing role, allowing particles to escape their original isopycnal layer and explore other stratified regions of the flow. Furthermore, this effect becomes more pronounced at lower Stokes numbers. From a dynamical perspective, this is expected: neutrally buoyant inertial particles act as band-pass filter, with a bandwidth proportional to $\textrm{St}$, making them in some cases more sensitive to turbulent fluctuations with certain frequencies \cite{Reartes2024}. Once again, for $\textrm{St} = 1$, the difference in vertical dispersion across subsets is smaller, consistent with enhanced particle clustering that inhibits vertical exploration. Thus, while inertial particles retain some degree of confinement and clustering, extreme events remain the primary mechanism capable of destabilizing their trajectories and enhancing vertical transport, in close analogy with the tracer dynamics.

\section{Conclusions}

In this work we explored in detail the relationship between extreme vertical velocity events in stratified turbulence and their impact on particle dispersion, focusing on the role of coherent structures, the dynamics of finite-time Lyapunov exponents (FTLEs), and energy dissipation. Using a Lagrangian approach, we showed that particle pair dispersion is strongly modulated by regions of the flow where intense dissipation occurs, establishing a clear connection between extreme events, coherent structures, and particle transport.

The analysis of tracer pair dispersion revealed that extreme events have a significant effect on vertical separation. Tracers experiencing such events exhibit markedly larger dispersion rates than those that do not. Moreover, in the TG simulation---where coherent structures are more prominent---dispersion is further amplified, suggesting that flow organization plays a central role in determining the Lagrangian transport efficiency. These results align with prior studies showing that stratified turbulence induces strong anisotropy in dispersion and enhances vertical mixing in localized regions \cite{Sujo_2018, Sujo_lag_2019, Feraco_2018}. The movie in the Supplementary Material \cite{sm} further emphasizes this result, providing a direct visualization of how extreme events can lead to dramatically different Lagrangian outcomes for tracers under identical flow conditions. Such events effectively reshape the dispersion landscape, acting as selective amplifiers of tracer separation.

The FTLE analysis confirmed that regions of high FTLE values coincide with the occurrence of extreme events, indicating that these flow structures are closely associated with strongly repelling Lagrangian trajectories. Rather than being uniformly distributed, extreme events cluster around dynamically organized regions that act as barriers or conduits for transport. The strong correlation between FTLEs and extreme events reinforces the idea that such regions not only enhance dispersion but also regulate intermittency in the transfer of energy across scales.

In stratified turbulence, coherent structures were found to modulate transport by organizing energy dissipation. In particular, the TG simulation displayed longer-lived structures with greater capacity to favour the emergence of extreme vertical drafts and to concentrate dissipation, as reflected in larger squared vorticity and stronger intermittency.
Our study of energy dissipation further showed that extreme events preferentially occur in regions of enhanced enstrophy, where the energy cascade culminates. Particles undergoing extreme events were found to exhibit significantly higher enstrophy than the rest, confirming that such events are intimately tied to localized zones of intense dissipation. This establishes a strong connection between dissipation and dispersion: the regions with the strongest energy dissipation coincide with those where particle separation is most pronounced, directly linking the energetic dynamics of the flow with Lagrangian transport. Finally, we showed that the ${\sim}20\%$ of tracers that intersect extreme vertical drafts account for nearly ${\sim}40\%$ of the total viscous dissipation sampled, underscoring the disproportionate energetic role of these localized events.
Additionally, we found that these high-dissipation events are intermittent and occur in well-defined patches of the flow, indicating that both transport and dissipation in stratified turbulence are not homogeneous processes but are dominated by localized zones where interactions between internal waves and turbulence generate extreme fluctuations. These findings are consistent with atmospheric observations, where clear-air turbulence (CAT) is seen to organize into sharply defined regions of strong dissipation, separated by quiescent zones \cite{rodriguezimazio2023}. The presence of distinct spectral structures within these events suggests that multiple mechanisms—possibly wave breaking, shear instabilities, and vortex stretching—modulate dissipation in each patch \cite{Marino2022, Smyth2013}.

In the case of neutrally buoyant inertial particles, our results reveal that extreme events significantly destabilize the vertical confinement imposed by stratification. Particles exposed to extreme events exhibit enhanced vertical dispersion, escape the isopycnal layer where they are otherwise stable, and explore a wider range of stratified layers. This effect is most pronounced for small Stokes numbers, where particles are more strongly coupled to the fluid and thus more responsive to fast vertical drafts. Finite-time Lyapunov exponents confirm that these particles follow highly repelling trajectories, leading to rapid separation at short times. At long times, while all particles tend to saturate at a vertical dispersion comparable to the buoyancy layer thickness, those experiencing extreme events consistently exceed the average, indicating a lasting imprint of extreme dynamics on their transport. These findings highlight that, even under neutral buoyancy, particle inertia plays a critical role in modulating the interaction with turbulent structures and in determining the effectiveness of vertical mixing in stratified flows.

This study has identified and quantified the relationship between extreme events, particle dispersion, energy dissipation, and flow structure in stratified turbulence. The results demonstrate that extreme events not only affect dispersion dynamics but also play a key role in the redistribution of energy throughout the system. These insights open pathways for improving the modelling of transport in geophysical flows, including applications to atmospheric mixing, pollutant dispersion, and the vertical transport of microplastics and biological material in the ocean \cite{Bourgoin2015, Riley2023}.

\begin{acknowledgments}
This work was supported by the projects Proyecto REMATE of Redes Federales de Alto Impacto, Argentina, and ``EVENTFUL'' (no.~ANR-20-CE30-0011) funded by the French Agence Nationale de la Recherche (ANR). The authors acknowledge HPC facilities at Departamento de Fisica (FCEN, UBA) in Argentina, and  at École Centrale de Lyon (PMCS2I) in Ecully (France), for the support in analyzing the data. CR also acknowledges the Universidad de Buenos Aires for funding a research trip in 2023, which made it possible to carry out this work.
\end{acknowledgments}

\bibliography{ms}

\begin{thebibliography}{68}
\bibitem [{sm()}]{sm}%
  \BibitemOpen
  \href@noop {} {}\bibinfo {note} {Side-by-side comparison of two tracer particle pairs evolving in the same stratified turbulent flow. The red pair crosses an intense vertical draft (visible in red), leading to rapid dispersion. The blue pair remains in a calmer region and stays closely bound. Both pairs are initialized with the same separation and evolve simultaneously in identical flow conditions, illustrating how extreme events can dramatically alter Lagrangian trajectories.}\BibitemShut {Stop}%
\end{thebibliography}

\end{document}